\documentclass[prb,twocolumn,showpacs,superscriptaddress,preprintnumbers,amssymb]{revtex4}
\usepackage{graphicx}
\usepackage{dcolumn}
\usepackage{bm}
\usepackage{wasysym}

\newcommand{\beq}{\begin{equation}}
\newcommand{\eeq}{\end{equation}}
\newcommand{\beqn}{\begin{eqnarray}}
\newcommand{\eeqn}{\end{eqnarray}}

\begin{document}
\title{Global phase diagram of the spin-1 antiferromagnet with uniaxial anisotropy on the kagome lattice}
\author{Cenke Xu}
\affiliation{Department of Physics, University of California,
Berkeley, CA 94720}
\author{Joel E. Moore}
\affiliation{Department of Physics, University of California,
Berkeley, CA 94720} \affiliation{Materials Sciences Division,
Lawrence Berkeley National Laboratory, Berkeley, CA 94720}
\date{\today}
\begin{abstract}
The phase diagram of the XXZ spin-1 quantum magnet on the kagome
lattice is studied for all cases where the $J_z$ coupling is
antiferromagnetic.  In the zero magnetic field case, the six
previously introduced phases, found using various methods, are:
the nondegenerate gapped photon phase which breaks no space
symmetry or spin symmetry; the six-fold degenerate phase with
plaquette order, which breaks both time reversal symmetry and
translational symmetry; the ``superfluid" (ferromagnetic) phase
with an in-plane global $U(1)$ symmetry broken, when $J_{xy} < 0$;
the $\sqrt{3}\times\sqrt{3}$ order when $J_{xy} > 0$; the nematic
phase when $D < 0$ and large; and a phase with resonating dimers
on each hexagon. We obtain all of these phases and partial
information about their quantum phase transitions in a single
framework by studying condensation of defects in the six-fold
plaquette phases.   The transition between nematic phase and the
six-fold degenerate plaquette phase is potentially an
unconventional second-order critical point. In the case of a
nonzero magnetic field along $\hat{z}$, another ordered phase with
translation symmetry broken is opened up in the nematic phase. Due
to the breaking of time-reversal symmetry by the field, a
supersolid phase emerges between the six-fold plaquette order and
the superfluid phase.  This phase diagram might be accessible in
nickel compounds, BF$_4$ salts, or optical lattices of atoms with
three degenerate states on every site.
\end{abstract}
 \maketitle

\section{introduction}

The behavior of ``frustrated'' magnets, in which not all
interaction energies can be simultaneously minimized, is already
quite complex when the individual spins are treated classically.
Models of quantum spins with frustrating interactions are an
active subject of current experimental and theoretical study.  A
simple example of a frustrated quantum magnet is the standard
nearest-neighbor Heisenberg antiferromagnet on any lattice with
closed loops containing an odd number of sites: important examples
include the triangular and the kagome lattices in two dimensions,
and the pyrochlore lattice in three dimensions.

For physical magnets with finite values of the spin $s$, there are
general approaches such as computing $1/s$ corrections to the
classical limit $s \rightarrow \infty$ and expanding the spin
algebra from $SU(2)$ to a larger group.  Such approaches are
powerful and predict many interesting ordered phases, but their
applicability to real magnets with only $SU(2)$ symmetry and small
values of the spin (e.g., $s=1/2$ or $s=1$) is uncertain.  In
recent years, interest has shifted to understanding specific
examples of finite-spin magnets in detail, even though the
necessary theoretical methods are less general than either the
$1/s$ or large-$N$ expansions.  Frustrated quantum
antiferromagnets with small spin $s=1/2$ or $s=1$ have been
proposed to show various exotic behaviors, including gapped or
algebraic spin liquids with gauge-boson-like excitations or
unconventional second-order phase
transitions~\cite{senthil2004,senthil2004a,ran2006}.

It is often possible to compare such predictions with large-scale
numerical Monte Carlo studies in cases with reduced symmetry
(e.g., with $SU(2)$ broken down to $U(1)$) , but frustrated
magnets with full $SU(2)$ symmetry are in general accessible only
by exact diagonalization, series expansion, or density-matrix
renormalization group on relatively small systems because of a
``sign problem'' associated with the frustration.  The $s=1$ model
on the kagome lattice studied in this paper is motivated both by
the existence of materials such as $\mathrm{BF_4}$
salts\cite{wada1997} and $\mathrm{Ni^{2+}}$-based materials
including $\mathrm{Ni_3V_2O_8}$\cite{lawes2004}, and by intrinsic
interest in the unexpected phases of the model.  Our goal is to
present a single treatment of the two-parameter phase diagram of
the model that unifies previous studies of parts of the phase
diagram~\cite{wen2003,xu2005,senthil2006} and allows consideration
of the various phase transitions occurring in the model.

Previous theoretical studies on the $s=1$ kagome lattice
antiferromagnet with uniaxial anisotropy (``XXZ anisotropy''),
with Hamiltonian \beq H = \sum_{<ij>}J_z S^z_iS^z_j + D(S^z_i)^2 +
J_{xy}(S_i^xS_j^x + S_i^yS_j^y),\label{hs1} \eeq Here the sum is
over nearest-neighbor bonds on the kagome lattice.  Note that the
on-site anisotropy term would be forbidden for $s=1/2$ and is
compatible with inversion symmetry, unlike the
Dzyaloshinksii-Moriya term, also quadratic in spin, that appears
if the other ions of the crystal break inversion symmetry.  For
general couplings, this Hamiltonian breaks the spin rotation
symmetry $SU(2)$ down to the $U(1)$ subgroup generated by $S^z$,
and has time-reversal symmetry.  We discuss both easy-plane and
easy-axis limits, and also consider briefly the effects of a
magnetic field that breaks time-reversal but preserves the $U(1)$.
Section II reviews previous theoretical work on the
zero-temperature physics of this Hamiltonian, which for different
values of the couplings has found a gapped phase with a massive
photon-like excitation~\cite{wen2003}, a critical line separating
plaquette-ordered phases~\cite{xu2005}, and an Ising-type spin
nematic~\cite{senthil2006}.  Section III presents the field-theory
description of the plaquette-ordered phases in terms of dual
height variables. From Section IV to section VII, we study the
transitions between the six-fold degenerate phase and other
phases, we will see that all the other phases can be interpreted
as the condensates of different kinds of defects in the six-fold
degenerate plaquette phases. In section VIII, the situation under
longitudinal magnetic field (along $\hat{z}$) is studied, several
new phases are found. Section IX is devoted to the point with
spin-$SU(2)$ symmetry, and section X is about other transitions in
phase diagram Fig. \ref{phasedia}.

\section{Experimental systems and previous studies}

So far two types of kagome spin-1 materials have been found. The
first type is $\mathrm{BF_4}$ salts \cite{wada1997}, the second is
$\mathrm{Ni^{2+}}$ based material $\mathrm{Ni_3V_2O_8}$
\cite{lawes2004}. Also, the kagome lattice has been constructed
with laser beams \cite{santos2004}, an effective spin model can
also be realized in cold atom system trapped in optical lattice,
but there the existence of biquadratic interactions comparable in
strength to the standard Heisenberg interaction makes the phase
diagram even more complicated \cite{demler2003}.

In general the model we are going to discuss is described by
equation (\ref{hs1}). This Hamiltonian is the simplest example
which can potentially realize all the physics discussed in this
work, but our formalism is supposed to be more general, and
independent of the details of the model on the lattice scale. This
is the simplest spin model which is invariant under time reversal
transformation. Three coefficients $J_z$, $J_{xy}$ and $D$ are
used to parameterize this model. If all the coefficients are
positive, this model can be realized in magnetic solids like those
given above; when $J_z, D > 0$, and $J_{xy} < 0$, this model could
possibly be realized in cold atom systems with pseudospin degrees
of freedom on each site. For instance, suppose on every site there
are three orbital levels (the three orbital levels can be the
degenerate $p$-level states, as discussed in several previous
papers \cite{girvin2005}), the orbital degrees of freedoms can be
viewed as spin-1 pseudospin, with natural XXZ symmetry. The
antiferomagnetic coupling $J_z$ and $D$ can be generated by the
on-site $s$-wave scattering and off-site dipole interactions
\cite{stuhler2005,axel2005}. The $J_{xy}$ coupling is resulted
from the superexchange, which should be ferromagnetic due to the
bosonic nature of the system. Therefore in the following
discussions, both positive $J_{xy}$ and negative $J_{xy}$ cases
will be discussed.

In solids, the spin $SU(2)$ symmetry can be broken by spin-orbit
coupling and the layered nature of the material, or by an external
magnetic field; in the cold atom pseudospin system, the $SU(2)$
symmetry is missing at the very beginning, as the orbital level
pseudospin system has natural uniaxial anisotropy.

Several previous papers have studied the kagome spin-1 system
\cite{wen2003,xu2005,senthil2006,levin2006}, at different
parameter regimes of this particular model (\ref{hs1}). There are
five phases that are already known.

{\bf Superfluid phase}: When $J_{xy} < 0$ and $|J_{xy}| \gg D,
J_z$, in this case $J_{xy}$ is the dominant term in the
Hamiltonian (\ref{hs1}): the expected phase is a superfluid phase
that breaks the global $U(1)$ symmetry. In the spin language this
phase is a ferromagnet in XY plane. Here the term superfluid phase
is used since the broken symmetry of this phase is the same as the
superfluid phase.

{\bf $\sqrt{3} \times \sqrt{3}$ phase}:  When $J_{xy} > 0$, and
becomes the dominant term in the Hamiltonian, the phase is not
obvious at first glance. When $D = J_z = 0$, and the spin $S
\rightarrow + \infty$, the system is at the classical XY limit. It
has been shown that the ground state of this classical XY model
has a large discrete degeneracy, in addition to a $U(1)$ that
rotates all the spins: the zero-temperature entropy associated
with this degeneracy is proportional to the size of the system.
The ground state configurations satisfy the requirement that every
triangle has zero net spin. If one spin is fixed, the whole ground
state configurations can be one to one mapped to the classical
ground states of the three-color model \cite{huse1992}.
Three-color model is defined as follows: on the honeycomb lattice,
each link is filled by one of the three colors, green, red and
blue, and the whole lattice is colored in such a way that every
site joins links of all three colors . The classical partition
function is defined as the equal weight summation of all the
3-color configurations. This partition function and entropy have
been calculated exactly by Baxter \cite{baxter1970}. It has also
been shown that the classical model can be mapped to a critical
2-component height model (similar to our model)
\cite{henley,read}, the low energy field theory of this model is a
$c = 2$ conformal field theory with $SU(3)_{k = 1}$ symmetry
\cite{kondev1996,read}.

\begin{figure}
\includegraphics[width=2.2in]{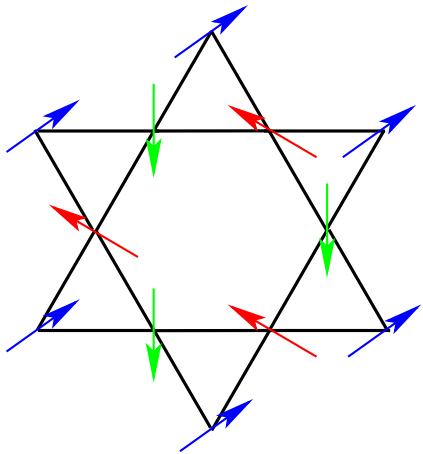}
\caption{the $\sqrt{3}\times\sqrt{3}$ order.} \label{root3}
\end{figure}

\begin{figure}
\includegraphics[width=2.2in]{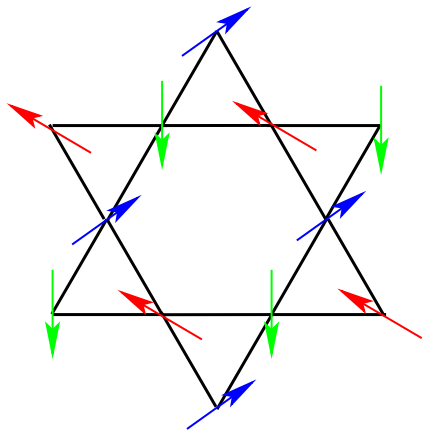}
\caption{the $q=0$ order.} \label{q=0}
\end{figure}

The large degeneracy of the classical model is not universal, and
it can be easily lifted by the second and third nearest neighbor
interaction $J_2$ and $J_3$. When $J_2 > J_3$, the $q = 0$ state
(Fig. \ref{q=0} ) is stabilized; while if $J_3 > J_2$, the
$\sqrt{3}\times\sqrt{3}$ state (Fig. \ref{root3}) is stabilized
\cite{harris1992}. The large 3-color degeneracy is also lifted by
$1/S$ expansion, and some ordered pattern is picked out from all
the classical degenerate ground states, this effect is usually
called ``order from disorder". At the isotropic case ($J_z =
J_{xy}$, $D = 0$), it was proved that after $1/S$ expansion both
coplanar $q = 0$ state and the $\sqrt{3}\times \sqrt{3}$ state are
stable \cite{chubukov1992}, i.e. they are both local minima in all
the ground states, the spin wave modes around these two minima do
not destabilize the order. Latter on, more detailed studies
suggest that the global minimum state is the $\sqrt{3}\times
\sqrt{3}$ order \cite{henley1995}, as depicted in Fig.
\ref{root3}. Although the $1/S$ expansion is carried out at the
isotropic point, the coplanar $\sqrt{3}\times \sqrt{3}$ phase is
expected to extend to the limit when $J_{xy}$ is dominant.

{\bf Gapped photon phase}: When $|J_{xy}| \ll J_z < D$, a gapped
phase without any symmetry breaking has been found \cite{wen2003}.
The low energy excitation with the smallest gap is a loop
excitation with the same polarization and gauge symmetry as a
photon: the effective theory can be described as a one-component
massive compact gauge field.

\begin{figure}
\includegraphics[width=2.5in]{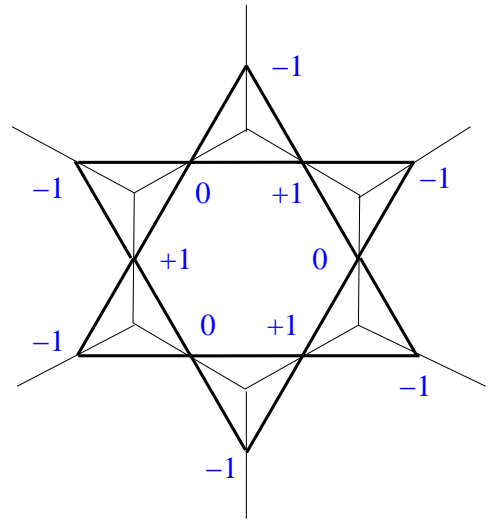}
\caption{The classical ground states of (\ref{hs1}), when $J_{xy}
= 0$. Every triangle has configuration $(1,-1,0)$, which can be
mapped to the 3 color model on the dual honeycomb lattice.}
\label{3color}
\end{figure}

{\bf Plaquette phase}: When $|J_{xy}| \ll D$ and $|J_z - D|$, $0 <
D < J_z$, a gapped phase with a six-fold degenerate ground state
has been found \cite{xu2005}. The six-fold degenerate ground state
has plaquette order: spins resonate around a subset of the
hexagons in the kagome lattice.  In this parameter regime, the
classical part of this model can be written as \beqn H =
\sum_{\triangle}\frac{J_z}{2}(\sum_{i = 1}^3S^z_i)^2 + \sum_{i}(D
- J_z)(S^z_i)^2. \eeqn When $0 < D < J_z$, the classical ground
states are all the configurations with every triangle occupied by
$S^z = (1,-1,0)$. Again the classical ground states can be mapped
onto the classical 3-color model \cite{baxter1970}, although the
3-color states correspond to $S^z$ instead of spins in the XY
plane (Fig. \ref{3color}).

\begin{figure}
\includegraphics[width=2.9in]{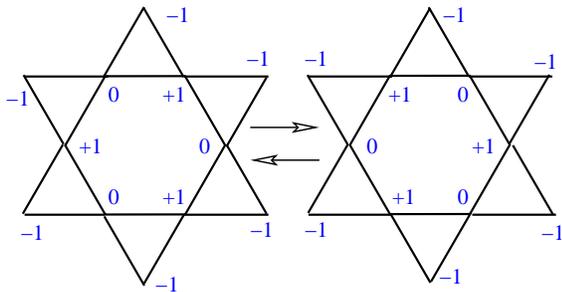}
\caption{The effect of the ring exchange term (\ref{ring}). It can
flip $\mathcal{A}_1$ ($\mathcal{B}_1$) to $\mathcal{A}_2$ ($
\mathcal{B}_2$), and vice versa.} \label{fring}
\end{figure}

If small $J_{xy}$ is turned on (either $J_{xy} > 0$ or $J_{xy} <
0$), the large degeneracy of 3-color ground states is lifted, and
the effective Hamiltonian which operates on the low energy Hilbert
space is \beqn H = \sum_{\hexagon} -t(S^\dagger_1 S^-_2S^\dagger_3
S^-_4S^\dagger_5 S^-_6 + H.c.). \label{ring}\eeqn 1 to 6 are the
sites of each hexagon on the kagome lattice. The flippable
hexagons have four kinds of configurations, they are
$(1,0,1,0,1,0)$ (denoted as $\mathcal{A}_1$), $(0,1,0,1,0,1)$
(denoted as $\mathcal{A}_2$), $(-1,0,-1,0,-1,0)$ (denoted as
$\mathcal{B}_1$) and $(0,-1,0,-1,0,-1)$ (denoted as
$\mathcal{B}_2$). The ring exchange term (\ref{ring}) can flip
$\mathcal{A}_1$ to $\mathcal{A}_2$ (and vice versa) Fig.
\ref{fring}, also can flip $\mathcal{B}_1$ to $\mathcal{B}_2$ (and
vice versa). Two compact $U(1)$ gauge fields were introduced to
describe this system, and due to the monopole proliferation, the
system is generally gapped, with crystalline order. The particular
order which happens here is the plaquette order, which breaks both
translational and time-reversal symmetries. The simplest way to
view this state is that, since the ring exchange term (\ref{ring})
can flip either $\mathcal{A}_1$ to $\mathcal{A}_2$ configurations,
or flip $\mathcal{B}_1$ to $\mathcal{B}_2$ configurations, the
configurations with the largest number of flippable hexagons are
favored in order to benefit from this ring exchange term. Notice
that the hexagons form a triangular lattice with three
sublattices, then one out of the three sublattices of the
triangular lattice can be resonated. Also one can choose either to
resonate between $\mathcal{A}_1$ and $\mathcal{A}_2$
configurations or to resonate between $\mathcal{B}_1$ and
$\mathcal{B}_2$ configurations (these two resonance cannot both
happen at the same state). Therefore there are in total $3\times 2
= 6$ degenerate ground states.

The simple picture of the ground state will be further justified
in the next section, by studying the dual quantum height model.
The classical height model was introduced to study the classical
3-color model, and since there are two components of free boson
height fields in the continuum limit, it is believed that the low
energy field theory should be $c = 2$ CFT \cite{kondev1996}. We
will see that the quantum effect is relevant at the classical
3-color critical point, a gap is opened due to the vertex
operators of the height fields.

Recently a mean-field treatment of a similar model has been
studied \cite{levin2006}. The plaquette phase we obtained is
similar but not entirely identical to the ``Plaquette ordered
phase" in this recent work, which is identified as the fully
packed string crystal. The main difference between the two
approaches is that, the monopole effect of compact gauge theory
has been taken into account in our work from the very beginning.
The monopole effect is supposed to be very relevant at the $z = 1$
Gaussian fixed point of gauge theory, and dominate the physics
close to the Gaussian fixed point. The nonlocal monopole effects
can be described by a local field theory in the dual formalism,
and the ordered pattern is predicted in this dual local field
theory.

\begin{figure}
\includegraphics[width=2.5in]{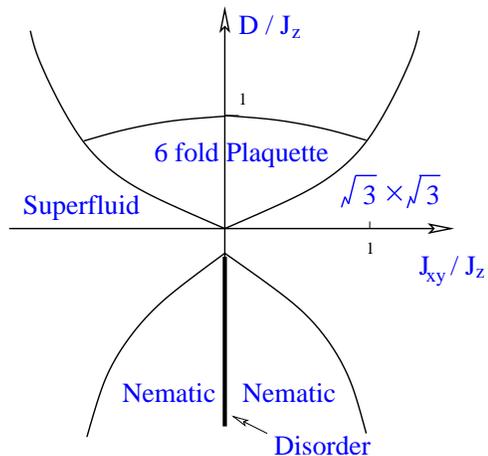}
\caption{the phase diagram with zero magnetic field. When $D >
J_z$ and $|J_{xy}|$ is small, the system is in the nondegenerate
gapped photon phase; when $0 < D < J_z$, and with small
$|J_{xy}|$, the ground state breaks both translational and time
reversal symmetry, resulting in the six-fold degenerate plaquette
order; when $J_{xy}$ is negative and large, basically the system
is in superfluid order which spontaneously breaks the global spin
$S^z$ conservation; when $J_{xy}$ is positive and large,
$\sqrt{3}\times\sqrt{3}$ order is  supposed to be favored; and
when $D$ is negative and large, the system has nematic order, with
nonzero expectation of $(S^+)^2$. } \label{phasedia}
\end{figure}

{\bf Nematic phase}: When $D < 0$ and $|J_{xy}|, J_z \ll |D|$, a
nematic phase with nonzero expectation value of $(S^\dagger)^2$
has been found \cite{senthil2006}. In this case, because $D$ is
negative and large, the system favors $S^z = \pm 1$ on every site.
Since the $S^z = 0$ state costs too much energy, every site can be
viewed as an Ising spin, and this model is effectively equivalent
to a spin-1/2 model.  Since $(S^-)^2$ flips $S^z = 1$ to $S^z =
-1$ state, it plays the same role as $\sigma^- = \sigma^x -
i\sigma^y$ on the effective Ising spin. Therefore the superfluid
phase of this spin-1/2 system is actually the nematic phase of the
original spin-1 model.

The rough phase diagram is shown in Fig. \ref{phasedia}. The goal
of the current work is to understand all the phases we know from
the excitations of the six-fold degenerate phase. Basically all
the phases can be interpreted as the condensates of various
defects which violate the $(1,-1,0)$ constraint in the plaquette
phase above. Since the low energy Hilbert space is a constrained
one, to create one single defect cannot be realized from local
moves of the ground state configurations, instead, global change
of all the spins is involved. This implies that one defect in this
phase not only carries the global $U(1)$ charge, but also carries
the gauge charge, with the gauge symmetry emerged at the low
energy constrained Hilbert space. Therefore the condensate of
defects is also the Higgs phase of the compact gauge fields.

\section{Gauge theory of the plaquette-ordered phase}

When $J_{xy} =0$ and $0 < D < J_z$, the set of degenerate ground
states can be mapped exactly~\cite{xu2005} to those of the 3-color
model~\cite{baxter1970}.  Every triangle on the kagome lattice has
$S^z$ configuration $(1, -1, 0)$ on this classical critical line.
The $z$-component spin configuration on the kagome lattice can be
viewed as two-component dimer configurations on the dual honeycomb
lattice, with repulsive interaction between two flavors of dimers
(every link of the honeycomb lattice can only be occupied by one
dimer). It is well-known that the one component quantum dimer
model can be mapped to compact gauge theory \cite{fradkin2004},
therefore it is natural to describe this spin-1 system as two
compact $U(1)$ gauge fields, since we can interpret the 3-color
constraint as two independent $U(1)$ constraints: every site on
the honeycomb lattice connects to exactly one $S^z = 1$ dimer and
one $S^z = -1$ dimer. We may map the three values of $S^z$ to
three configurations of a two-component electric field:

\begin{eqnarray}
S^z_i = 0 \Rightarrow (E_1, E_2) = \frac{1}{\sqrt{3}}(1, 0),\cr
S^z_i = 1 \Rightarrow (E_1, E_2) = \frac{1}{\sqrt{3}}( -
\frac{1}{2}, \frac{\sqrt{3}}{2}),\cr S^z_i = -1 \Rightarrow (E_1,
E_2) = \frac{1}{\sqrt{3}}( - \frac{1}{2}, - \frac{\sqrt{3}}{2}
).\label{e}
\end{eqnarray}

Next, note that a 2D unit vector $\hat{\mathbf{n}}_i$ can be
assigned parallel or antiparallel to each link $i$ of the
honeycomb lattice (dual lattice of the kagome lattice) so that
vertices of sublattice $A$ of the honeycomb have three incoming
bonds, while those of sublattice $B$ have three outgoing bonds.
Now define two-component vector fields on bonds: $\vec{E}_\alpha =
E_\alpha\hat{\mathbf{n}}_i$. The three color constraint is now
equivalent to the Gauss's law constraint

\begin{eqnarray}
\vec{\nabla}\cdot \vec{E}_1 = \vec{\nabla}\cdot \vec{E}_2 = 0.
\label{gaussian}
\end{eqnarray}

Also we can generalize the configuration of the $E$ vector to a 2d
triangular lattice. The lattice is formed by basis $\vec{b}_1 =
(\sqrt{3}/2, 1/2)$ and $\vec{b}_2 = (0, 1)$, \beqn \vec{E} =
n\vec{b}_1 + m\vec{b}_2 - (1/(2\sqrt{3}),1/2) \eeqn If we add the
following interaction to the Hamiltonian, the $\vec{E}$ fields
only take three smallest vectors as (\ref{e}):

\begin{eqnarray}
\frac{1}{\kappa}(\vec{E}_1^2 + \vec{E}_2^2).
\end{eqnarray}

Thus the low energy configurations of electric fields can be
one-to-one mapped to the low energy configurations of spins; the
spin formalism and the electric field formalism are equivalent.

The perturbation theory of $J_{xy}$ generates a ring exchange
(\ref{ring}). The ring-exchange term breaks the $\mathbb{Z}_3$
symmetry. Define conjugate operators on each bond $(A_{1,i},
A_{2,i})$ with commutation relations \beqn
[A_{\alpha,j},E_{\beta,k}] = i\delta_{\alpha\beta}\delta_{kj}.
\eeqn Then operator $T_{\alpha,i} = \exp(iA_{\alpha,j})$ acts as a
raising operator: it increases the quantum number $E_{\alpha,j}$
by 1. This enables a compact representation of the ring-exchange
terms proportional to $t$: on bond $j$,
$\exp(iA_{\alpha,j}l^{(1)}_\alpha)$ will raise $S^z_{j} = 0$ to
$S^z_{j} = 1$ if $l^{(1)} = (-\sqrt{3}/2, 1/2)$. Similarly, if
$l^{(2)} = (-\sqrt{3}/2, -1/2)$ then
$\exp(iA_{\alpha,j}l^{(2)}_\alpha)$ takes $S_j^z = 0$ to $S_j^z =
-1$. Define vector $\vec{A}_{\alpha,j} =
A_{\alpha,j}\hat{\mathbf{n}}_j$, the ring exchange term around
each hexagon becomes

\begin{eqnarray}
H_{ring} = \sum_{i = 1}^2-t\cos(\vec{\nabla}\times\vec{A}_\alpha
l^{(i)}_\alpha).\label{ringgauge}
\end{eqnarray}

Here as usual in gauge theories of lattice spin models, the
meaning of $\vec{\nabla} \times \vec{A}$ is that one takes the
lattice circulation around a plaquette: for $\hat{v}_j$ an
clockwise assignment of unit vectors along the links around a
hexagon \beqn \vec{\nabla}\times \vec{A} = \sum_{j = 1}^6
\hat{v}_j\cdot \vec{A}_j .\eeqn

If no defect is present, i.e. the Gauss's law constraint is
strictly imposed, the theory is described by two compact $U(1)$
gauge fields without matter fields. Now let us consider the
defects, which are also the gauge charges. When $D$ is much
smaller than $J_z - D$, the excitation with the smallest gap is to
flip one site with $S^z = 0$ to 1 (or -1), this process actually
creates a pair of $(1,1,-1)$ (or $(-1,-1,1)$) defects. Let us
denote the density of the $(1,1,-1)$ configuration defect as
$\rho_1$, and denote the density of the $(1,-1,-1)$ defect as
$\rho_2$, then from the definition of electric field we can obtain
the following relations

\begin{eqnarray}
\vec{\nabla}\cdot \vec{E}_1 = - \frac{\sqrt{3}}{2}(\rho_1 +
\rho_2)\cr \vec{\nabla}\cdot \vec{E}_2 = \frac{1}{2}(\rho_1 -
\rho_2).\label{defect}
\end{eqnarray}

The charges can be effectively viewed as matter fields defined on
the sites of the honeycomb lattice, and the gauge fields $\vec{A}$
and $\vec{E}$ are fields defined on the links of the honeycomb
lattice.

For the convenience of later calculations, we need to define a new
set of variables as follows

\begin{eqnarray}
\vec{e}_1 = - \frac{1}{\sqrt{3}}\vec{E}_1 + \vec{E}_2 \cr
\vec{e}_2 = - \frac{1}{\sqrt{3}}\vec{E}_1 - \vec{E}_2 \cr
\vec{a}_1 = - \frac{\sqrt{3}}{2} \vec{A}_1 + \frac{1}{2}\vec{A}_2
\cr \vec{a}_2 = - \frac{\sqrt{3}}{2} \vec{A}_1 -
\frac{1}{2}\vec{A}_2.\label{ea}
\end{eqnarray}

Also, one can check that $\vec{e}_\alpha$ and $\vec{a}_\alpha$ are
still conjugate variables :

\begin{eqnarray}
[e_{\alpha,i}, a_{\beta,j}] = i\delta_{ij}\delta_{\alpha\beta}.
\end{eqnarray}

If the definition for $\vec{e}_\alpha$ and $\vec{a}_\beta$ is
plugged in (\ref{defect}), one can see that $\vec{e}_\alpha$ is
the electric field corresponding to the charge $\rho_\alpha$, in
the sense that

\begin{eqnarray}
\vec{\nabla}\cdot \vec{e}_\alpha = - \rho_\alpha.
\label{gaugecharge}
\end{eqnarray}

When $D$ is smaller than but close to $J_z$, $|J_z - D| \ll D$,
the lowest energy excitation is $(0,0,0)$, and we denote its
density as $\rho_0$.  It carries gauge charge of gauge field
$\vec{E}_1$

\begin{eqnarray}
\vec{\nabla}\cdot \vec{E}_1 = \sqrt{3}\rho_0 = -
\frac{\sqrt{3}}{2}(\vec{\nabla}\cdot \vec{e}_1 + \vec{\nabla}\cdot
\vec{e}_2)
\end{eqnarray}

Since the electric fields are subject to the constraint
(\ref{gaussian}), it is convenient to define height fields
$\vec{h}$ on the dual triangular lattice.

\begin{eqnarray}
\vec{E}_\alpha = (\hat{z}\times\vec{\nabla}) h_\alpha,
\vec{\nabla}\times \vec{A}_\alpha = \pi_{h\alpha}.\label{height}
\end{eqnarray}

$\pi_{h\alpha}$ and $h_\alpha$ are a pair of conjugate variables.
The value of $h_\alpha$ is also defined on a triangular lattice
configuration space, in order to satisfy definition
(\ref{height}), $h_\alpha$ is defined in the following way

\begin{eqnarray}
(h_1, h_2)_a = (\frac{\sqrt{3}}{2}(m + n), \frac{1}{2}(m - n)) +
(q_1, q_2)_a,
\end{eqnarray} $m$ and $n$ are both integers. Here $a = A, B, C$, denoting the
three sublattices on the triangular lattice (dual lattice of the
honeycomb lattice). $\vec{q}_a$ are three vectors, taking
different values on three sublattices

\begin{eqnarray}
\vec{q}_A = (\frac{1}{2\sqrt{3}}, - \frac{1}{6})\cr \vec{q}_B =
(-\frac{1}{2\sqrt{3}}, - \frac{1}{6})\cr \vec{q}_C = (0 ,
\frac{1}{3})
\end{eqnarray}

The two-component height variables $h_\alpha$ are the same as
those introduced in the classical 3-color model (cf. Kondev and
Henley \cite{kondev1996}). Since $m$ and $n$ are both integers,
the vertex operators should enter the effective low energy theory.
We will see later that, due to quantum effect, these vertex
operators become very relevant and drive the system away from the
classical criticality, resulting in a six-fold degenerate
plaquette ordered phase. These vertex operators read

\begin{eqnarray}
H_{v} = \sum_{a = A}^C-\alpha\cos[2\pi(\frac{1}{\sqrt{3}}h_{1a} +
h_{2a} + \frac{1}{\sqrt{3}}q_{1a} + q_{2a}) ]\cr - \sum_{a = A}^C
\alpha\cos[2\pi(\frac{1}{\sqrt{3}}h_{1a} - h_{2a} +
\frac{1}{\sqrt{3}}q_{1a} - q_{2a}) ].
\end{eqnarray}

For later convenience, we define a new height fields $\phi_{ia}$
and its conjugate variable $\pi_{ia}$ as

\begin{eqnarray}
\phi_{1} = \frac{1}{\sqrt{3}}h_{1} + h_{2} \cr \phi_{2} =
\frac{1}{\sqrt{3}}h_{1} - h_{2} \cr \pi_{1} =
\frac{\sqrt{3}}{2}\pi_{h1} + \frac{1}{2}\pi_{h2}\cr\pi_{2} =
\frac{\sqrt{3}}{2}\pi_{h1} - \frac{1}{2}\pi_{h2}
\end{eqnarray}

One can check the commutators and see that $\phi_{\alpha}$ and
$\pi_{\alpha}$ are conjugate variables, and based on the
definition (\ref{ea}), they are exactly the height fields
corresponding to $\vec{e}$ and $\vec{a}$.

\begin{eqnarray}
\vec{e}_\alpha = (\hat{z}\times \vec{\nabla}) \phi_\alpha,
\vec{\nabla} \times \vec{a}_\alpha = \pi_\alpha.
\end{eqnarray} The vortex of $\phi_\alpha$ is the charge field
$\rho_\alpha$.

Now in terms of the new height fields, the vertex operators read

\begin{eqnarray}
H_v =  -\alpha[\cos(2\pi\phi_{1A}) + \cos(2\pi\phi_{1B} +
\frac{2\pi}{3}) \cr + \cos(2\pi\phi_{1C} + \frac{4\pi}{3}) \cr +
\cos(2\pi\phi_{2A} + \frac{4\pi}{3}) + \cos(2\pi\phi_{2B} ) \cr +
\cos(2\pi\phi_{2C} + \frac{2\pi}{3})].\label{vertex1} \cr
\end{eqnarray}

These vertex operators have oscillating signs on the triangular
lattice, then in the low energy theory the relevant terms should
be higher orders of vertex operators which do not contain
oscillating signs on the lattice:

\beqn H_v = -v[\cos(6\pi\varphi_1) + \cos(6\pi\varphi_2)] \cr\cr -
v_1[\cos(2\pi\varphi_1 + 4\pi\varphi_2) + \cos(4\pi\varphi_1 +
2\pi\varphi_2) ]\cr\cr - v_2\cos(2\pi\varphi_1 - 2\pi\varphi_2) +
\cdots\label{vertex2}\cr \eeqn

$\varphi_\alpha$ is the coarse-grained mode of $\phi_\alpha$. As
the vertex operators correspond to the creation and annihilation
of gauge fluxes, the total gauge flux is conserved by mod 3 in the
low energy continuum limit. In equation (\ref{vertex2}), $v$ is
supposed to be positive, but $v_1$ and $v_2$ are supposed to be
negative, because when we subtract $\phi_2$ from $\phi_1$ from
(\ref{vertex1}), it gains angle $2\pi/3$, which generates a factor
$-1/2$ before the cosine term in (\ref{vertex2}). Sine functions
of $\varphi_\alpha$ are excluded by symmetries of the system. For
instance, $\sin(2\pi(\varphi_1 - \varphi_2))$ is excluded by time
reversal symmetry.

After coarse-graining the system, the action in terms of
$\varphi_\alpha$ can be written as \beqn L = \sum_{\alpha =
1}^2(\partial_\tau \varphi_\alpha)^2 +
\rho_2(\nabla\varphi_\alpha)^2 + H_v + \gamma
\nabla\varphi_1\nabla\varphi_2 \label{heighth}\eeqn

The $\gamma$ term in (\ref{heighth}) is a flavor mixing term
between $\varphi_1$ and $\varphi_2$, and therefore the two flavors
of height fields do not only couple to each other through the
vertex operators, but also through one of the kinetic terms.

In 2+1d, the potential operators with cosine functions are
generally very relevant at the fixed point described by the
Gaussian part of the action (\ref{heighth}), as long as the $k^2$
term ($\rho_2$ in (\ref{heighth})) is present. Vertex operators
are responsible for the gapped crystalline phases of quantum dimer
models, both on the square lattice \cite{kivelson1988,fradkin1990}
and the honeycomb lattice \cite{fradkin2004}. In the current work,
the vertex operators are also responsible for the crystalline
phases. First of all, let us tune $v_1$ and $v_2$ to zero, and
minimize $v$ terms in (\ref{vertex2}). Each $\varphi_\alpha$ has
three minima $0$ , $1/3$, $2/3$. Therefore there are in total 9
different combinations. However, negative $v_1$ and $v_2$ terms
will raise the energy of all the minima with $\varphi_1 =
\varphi_2$, and hence we end up with $9 - 3 = 6$ minima. This
result is actually quite general, for a large parameter regime,
there are always 6 minima of the vertex potential in
(\ref{vertex2}). Because the vertex operators in (\ref{vertex2})
is invariant under transformation $\varphi_1 \rightarrow \varphi_1
+ 1/3$, $ \varphi_2 \rightarrow \varphi_2 + 1/3$, and also
invariant under transformation $\varphi_1 \rightleftarrows
\varphi_2$, all six minima can be obtained from performing
transformations on one single minimum.

Now we can write down the plaquette order parameter in terms of
the field theory variables $\varphi_\alpha$. The order parameter
we are searching for, in the lattice model, is \beq P = \sum_{a =
A}^C e^{i2\pi(i_a-1)/3}(S^\dagger_{1a} S^-_{2a}S^\dagger_{3a}
S^-_{4a}S^\dagger_{5a} S^-_{6a} + H.c.)(\sum_{\hexagon}S^z_i) \eeq

In the above equation, $a$ represents different sublattices, and
$i_A = 1$, $i_B = 2$, and $i_C = 3$. The low energy representation
of this order parameter can be deduced from symmetry argument. The
most obvious transformations for this order parameter are
translational ($\mathrm{T}$) and time reversal ($\mathrm{TR}$)
transformations. \beq \mathrm{T} : P \rightarrow e^{i2\pi/3}P,
\mathrm{TR}: P \rightarrow - P. \eeq

If rotated around hexagons at sublattice $A$ by angle $2\pi/3$
($\mathrm{R_{2\pi/3}}$), the order parameter is invariant; under
space inversion (SI) $\vec{r}\rightarrow -\vec{r}$ and reflection
($\mathrm{P_x}$) along $\hat{x}$ ( $y \rightarrow -y$ ) centered
at sublattice $A$, the order parameter becomes its complex
conjugate\beqn \mathrm{SI}, \mathrm{P_x}: P \rightarrow P^\ast.
\eeqn

Under the transformations discussed above, $\varphi_i$ transforms
as follows: \beqn \mathrm{T} : \varphi_i \rightarrow \varphi_i +
\frac{1}{3}, \cr \mathrm{TR} : \varphi_1 \rightleftharpoons
\varphi_2, \cr \mathrm{SI} : \varphi_\alpha \rightarrow -
\varphi_\alpha, \cr \mathrm{P_x}: \varphi_\alpha \rightarrow -
\varphi_\alpha, \cr \mathrm{R_{2\pi/3}}: \varphi_i \rightarrow
\varphi_i. \eeqn

Summarizing all the transformations above, the field theory
representation of the plaquette order parameter $P$ is \beqn P
\sim (e^{i2\pi\varphi_2} - e^{i2\pi\varphi_1}).\label{plafield}
\eeqn

We can plug in the six minima of the vertex operator
(\ref{vertex2}) to (\ref{plafield}), and it gives us 6 different
values. All the six expectation values can be obtained by
transformation $\langle P_n \rangle = \exp(i\pi n/3)\langle
P_0\rangle$, with $n = 1$ to $6$. This implies that the system is
in a plaquette order with six fold degeneracy.

When both $v_1$ and $v_2$ are positive, the vertex operators gives
three degenerate minima: $\varphi_1 = \varphi_2 = 0$, or $\pm
1/3$. These three degenerate ground states do not break time
reversal symmetry, but it breaks translational symmetry. The
particular order in this case is another type of plaquette order
with $(-1,1,-1,1,-1,1)$ hexagons resonating on one of the three
sublattices.

The field theory description is only valid when the theory is
close to a critical point, i.e. the correlation length is either
infinite, or finite but much longer than the microscopic lattice
constant. Thus the prediction of plaquette order is only rigorous
close to the classical critical point $J_{xy} = 0$. But the phase
is expected to extend over a finite region in the phase diagram,
until a transition into either a disordered phase or one of the
other ordered phases derived in the following sections.

In this section we started with mapping the classical ground
states of the model onto the classical 3-color model
configurations, as in this way we respected the $\mathbb{Z}_3$
symmetry of the classical ground state, which is broken by the
quantum perturbation. As mentioned before, we can also view the
low energy physics of this system as two components of quantum
dimer model, with repulsive interaction between two flavors of
dimers. From this approach the same low energy action as
(\ref{heighth}) can be derived. Single component of quantum dimer
model generates the kinetic terms and the vertex operator
$\cos(6\pi\varphi_i)$ in (\ref{heighth}) at low energy, as
discussed in reference \cite{fradkin2004}; the repulsive
interaction between the two flavors of dimers will generate the
term $\gamma \nabla\varphi_1\nabla\varphi_2$ and the mixture
vertex operators $\cos(2\pi(\varphi_1 - \varphi_2))$ and
$\cos(2\pi\varphi_1 + 4\pi \varphi_2)$, et,al.

\section{transition to the featureless gapped photon phase}

When $D > J_z$, the classical ground state is $S^z = 0$ on every
site, and the low energy excitations are $(+1,-1,+1,-1,....)$
loops. This phase has a single ground state and gapped photon
excitations \cite{wen2003}, without any symmetry breaking. In this
section we are going to study the phase transition between the
six-fold state and the gapped photon phase.

If we start with the six-fold degenerate phase, the transition can
be viewed as condensation of $(0,0,0)$ defects. The gap for
$(0,0,0)$ defect keeps decreasing as the transition to the gapped
photon phase is approached. But the phase boundary between the
plaquette phase and the nondegenerate phase is not exactly at $D =
J_z$ (Fig. \ref{phasedia}), this is due to the fact that at second
order perturbation of $J_{xy}/J_z$, an additional nearest neighbor
diagonal interaction is generated. $(0,0,0)$ triangles are more
favorable than $(1,-1,0)$ triangles to this diagonal term
generated, therefore the second order perturbation effectively
increases $D$ by $\sim J_{xy}^2/J_z$.

The defect $(0,0,0)$ carries charges of both $a_{1\mu}$ and
$a_{2\mu}$, and defects at different sublattices of the honeycomb
lattice carry opposite gauge charges. If we denote the $(0,0,0)$
defect at sublattice $A$ as $\psi_A$ and the $(0,0,0)$ defect at
sublattice $B$ as $\psi_B$, the effective Lagrangian describing
the system close to the transition is \beqn L = - t|(\partial_\mu
- ia_{1\mu} - ia_{2\mu})\psi_{A}|^2 - t|(\partial_\mu + ia_{1\mu}
+ ia_{2\mu})\psi_{B}|^2 \cr \label{000action}\eeqn Notice that the
$(0,0,0)$ defect carries zero global $U(1)$ charge (a $(0,0,0)$
defect does not carry any $S^z$), and therefore one $\psi_A$
particle and one $\psi_B$ particles can be annihilated together,
so the term $g(\psi_A\psi_B + H.c.)$ is allowed in the
interaction. After the condensation of $\psi_A$ and $\psi_B$, the
gauge field $a_{+\mu} = a_{1\mu} + a_{2\mu}$ will be gapped out
along with the phase mode $\theta_A - \theta_B$, and the mode
$\theta_A + \theta_B$ will be gapped out by the interaction
$g(\psi_A\psi_B + H.c.)$ ($\theta_A$ and $\theta_B$ are phase
angles of $\psi_A$ and $\psi_B$ respectively). Therefore in the
condensate there is no gapless excitation, which is consistent
with the gapped photon phase.

To further justify this picture, let us first take a
Landau-Ginzburg tour to study this transition. Let us define
complex field $\Phi$ to describe the low energy mode of the
plaquette order parameter $P$. The LG action for this transition
is as follows \beqn L = |\partial_\mu\Phi|^2 - r|\Phi|^2 +
u(|\Phi|^2)^2 + g(\Phi^6 + H.c.). \label{LG}\eeqn Without the $g$
term, the theory describes an 3D XY transition. The $g$ term turns
on an $\mathbb{Z}_6$ anisotropy at this critical point. In the
ordered state of $\Phi$, this anisotropy is a relevant
perturbation and will lead to a six-fold degeneracy. At the $3D$
XY critical point, $\mathbb{Z}_6$ anisotropy is irrelevant, thus
the Landau-Ginzburg theory predicts that the transition between
the six-fold states and the featureless gapped photon phase is a
3D XY transition.

\begin{figure}
\includegraphics[width=3.3in]{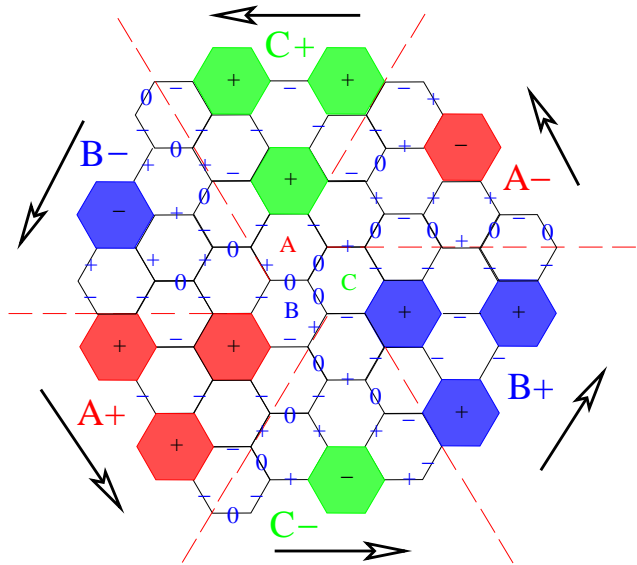}
\caption{The vortex and vortex core configuration in the six-fold
plaquette order. Every vortex is surrounded by six phase domains
and 6 domain walls, the vortex core is exactly a $(0,0,0)$ defect.
In this figure, $\mathrm{B+}$ denotes resonating $(1,0,1,0,1,0)$
plaquette on sublattice $\mathrm{B}$. The six domains around this
vortex core are (count counterclockwise): $A+$ with $\Phi \sim
\Phi_0$, and $\vec{\varphi} = \vec{\varphi}_0 + (1/3, 2/3)$; $C-$
with $\Phi \sim \exp(i\pi/3)\Phi_0$ and $\vec{\varphi} =
\vec{\varphi}_0 + (1/3, 0)$, $B+$ with $\Phi \sim
\exp(i2\pi/3)\Phi_0$ and $\vec{\varphi} = \vec{\varphi}_0 + (2/3,
0)$; $A-$ with $\Phi \sim \exp(i\pi)\Phi_0$ and $\vec{\varphi} =
\vec{\varphi}_0 + (2/3, 1/3)$; $C+$ with $\Phi \sim
\exp(i4\pi/3)\Phi_0$ and $\vec{\varphi} = \vec{\varphi}_0 + (0,
1/3)$, $B-$ with $\Phi\sim\exp(i5\pi/3)\Phi_0$ and $\vec{\varphi}
= \vec{\varphi}_0 + (0, 2/3)$. The arrows show the circulation of
the phase of $\Phi$. Also, this vortex is a bound state of one
vortex of $\varphi_1$ and one vortex of $\varphi_2$. The spins are
defined on the links of the honeycomb lattice, which are the sites
of the kagome lattice. links with + and - are occupied by spin
states $S^z = +1$ and $-1$. The red dashed lines denote the domain
walls.} \label{core}
\end{figure}

The 3D XY transition is driven by the vortices of $\Phi$, and
after the condensation of the vortices, the vortex core state
grows and becomes the macroscopic order. It has been shown before
that the vortex core of the height field of the quantum dimer
model is an unpaired spin, which implies that the condensate of
vortices breaks the spin $SU(2)$ symmetry spontaneously (for
instance, the Neel state). In our case, the vortex configuration
of $\Phi$ (including the core) has been depicted in Fig.
\ref{core}. Around every vortex core, there are 6 domains
separated by domain walls, each domain is one state out of the
six-fold degenerate plaquette ordered states. In the ordered
phase, the vortices are linearly confined due to the pinning
potential $\Phi^6 + H.c.$, because the domain walls would cost
energy proportional to their length. At the critical point since
the pinning potential is irrelevant, the vortices are deconfined.

In Fig. \ref{core}, one can see that the vortex core is actually a
(0,0,0) triangle, which is the lowest energy defect when $D \sim
J_z$. If the height field representation of $\Phi$
(\ref{plafield}) is taken, one can see that the vortex of $\Phi$
is a bound state of one vortex of $\varphi_1$ and one vortex of
$\varphi_2$. Thus a vortex of $\Phi$ carries one gauge charge of
$a_{1\mu}$ and one gauge charge of $a_{2\mu}$, i.e. this vortex
carries the same gauge charge as the $(0,0,0)$ defect. Therefore
indeed the transition between the six-fold plaquette state and the
featureless photon phase is driven by the $(0,0,0)$ defects.

The dual field theory of (\ref{LG}) would describe the vortex
condensation directly. After the standard superfluid-gauge field
duality in 2+1d, the dual theory reads \beqn L = -t|(\partial_\mu
- iA_\mu )\psi|^2 + \cdots \label{LGdual}\eeqn Herein $\psi$ is
the vortex creation operator, and the $\mathbb{Z}_6$ anisotropy
term in (\ref{LG}) becomes the monopole processes which annihilate
and create the fluxes of gauge field $A_\mu$. By comparing
equation (\ref{LGdual}) and equation (\ref{000action}), we can see
that $A_\mu = a_{1\mu} + a_{2\mu}$, and $\psi = \psi_A =
\psi_B^\dagger$. Please note that because $\psi_A$ and $\psi_B$
can annihilate together, there is actually only one flavor of
defect, and $\psi_B = \psi_A^\dagger$. In the ordered phase the
gauge field $A_\mu$ is gapped out by monopole proliferation, and
$\psi$ is confined; in the nondegenerate photon phase the gauge
field is gapped out with $\psi$ through the Higgs mechanism. The
gauge field is only gapless at the critical point.

If we plug in the height field representation of $\Phi$ in
equation (\ref{plafield}) into the LG action (\ref{LG}), it
reproduces the height field action in equation (\ref{heighth}),
thus the phase transition between the gapped photon phase and the
plaquette phase can also be studied in the dual height model. We
define new height fields $\varphi_{\pm} =
(\varphi_1\pm\varphi_2)/2$, and they satisfy the following
relation \beq (\vec{e}_{1} \pm \vec{e}_2)/2 =
(\hat{z}\times\vec{\nabla}) \varphi_{\pm}. \eeq Now the height
field Lagrangian reads \beqn L = t(\partial_\tau\varphi_+)^2 +
\rho_2(\nabla\varphi_+)^2 + t^\prime(\partial_\tau\varphi_-)^2 +
\rho^\prime_{2}(\nabla\varphi_-)^2 \cr\cr -v
\cos(6\pi\varphi_+)\cos(6\pi\varphi_-) - v_1
\cos(6\pi\varphi_+)\cos(2\pi\varphi_- )\cr\cr - v_2
\cos(4\pi\varphi_-). \label{000}\eeqn

One $(0,0,0)$ defect carries one unit gauge charge of $(e_1 +
e_2)/2$, thus it is one unit vortex of $\varphi_+$, and the
condensation of $(0,0,0)$ drives $\varphi_+$ into disordered
phase. In the condensate, $\varphi_+$ is disordered and the
expectation value of $\cos(2\pi\varphi_+)$ is zero. Thus the
plaquette order parameter \beqn P &\sim& e^{i2\pi \varphi_2} -
e^{i2\pi \varphi_1} \sim i \sin(2\pi\varphi_-)
\exp(i2\pi\varphi_+) \eeqn takes zero expectation value: the
plaquette order disappears. Since $\varphi_-$ does not transform
under translation or rotation by $2\pi/3$ transformations, any
crystalline pattern which breaks these symmetries cannot exist.

When field $\varphi_+$ is disordered, the ordered pattern and
symmetry of the ground state can be studied from the effective
action for height field $\varphi_-$, which remains gapped and
ordered. Thus the order of the condensate is determined by the
series of vertex operators of $\varphi_-$, since the leading
vertex operator is $-v_2\cos(4\pi\varphi_-)$, for a large range of
parameters, the minima are at $\varphi_- = \pm 1/4$. However, let
us imagine writing down a physical order parameter which only
involves $\varphi_- = (\varphi_1 - \varphi_2)/2$, since any
physical order parameter should be invariant under transformation
$\varphi_\alpha \rightarrow \varphi_\alpha + 1$, this order
parameter should be invariant under $\varphi_- \rightarrow
\varphi_- + 1/2$, thus the ground states $\varphi_- = \pm 1/4$ are
physically equivalent to each other, and the ground state is
nondegenerate, which is again consistent with the gapped photon
phase.

At the transition, since only $\varphi_-$ is ordered, we can plug
in the minimum of $\varphi_-$ into (\ref{000}), and obtain an
effective action for $\varphi_+$. Notice that both $v$ and $v_1$
vertex operators vanish after plugging in the minima $\varphi_- =
1/4$. The leading operator that survives is
$\cos(12\pi\varphi_+)$, which is a $\mathbb{Z}_6$ anisotropy. The
height field theory which describes this transition is \beqn L =
(\partial_\mu\varphi_+)^2 - \alpha \cos(12\pi\varphi_+). \eeqn
This action describes an XY transition as the $\mathbb{Z}_6$
anisotropy term is irrelevant at the XY critical point. Thus we
conclude that the transition between the six-fold state and the
featureless photon phase is driven by the condensation of
$(0,0,0)$ defect, and the critical point belongs to the 3D XY
universality class.

\section{transition to the superfluid state}

\begin{figure}
\includegraphics[width=3.3in]{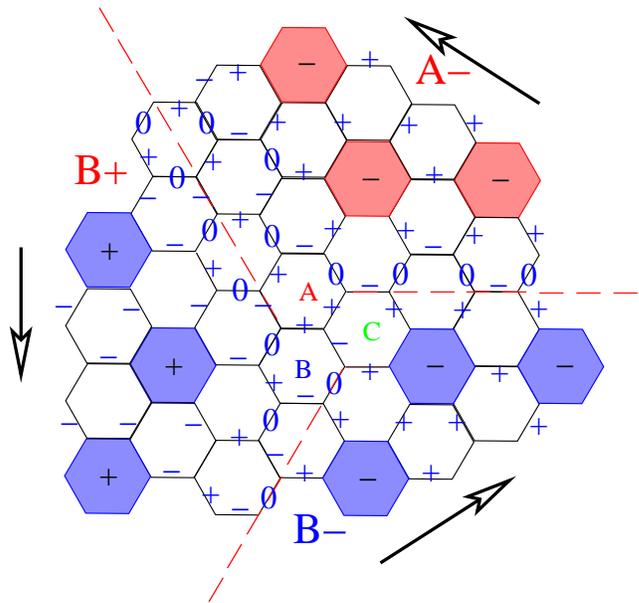}
\caption{The vortex and vortex core configuration of $\varphi_1$.
Every vortex is surrounded by 3 phase domains and 3 domain walls,
the vortex core is exactly a $(1,1,-1)$ defect. The three domains
around this vortex core are (count counterclockwise): $B+$ with
$\vec{\varphi} = \vec{\varphi}_0 + (2/3, 0)$; $B-$ with
$\vec{\varphi} = \vec{\varphi}_0 + (0, 2/3)$, $A+$ with
$\vec{\varphi} = \vec{\varphi}_0 + (1/3, 0)$. The arrows show the
circulation of $\varphi_1$. } \label{core3}
\end{figure}

When $ J_{xy}$ is negative and large, the system is in the
superfluid phase (ferromagnetic phase in spin XY plane), with
nonzero expectation value of $\langle S^\dagger \rangle$. In this
section we are going to study the transition between the six-fold
degenerate plaquette phase and the superfluid phase. Let us first
focus on the region where $D \ll |J_z - D|$; in this parameter
regime, the defects with the lowest gap are $(1,1,-1)$ and
$(1,-1,-1)$ triangles. It was shown in section III that these two
defects are vortices of $\varphi_1$ and $\varphi_2$ respectively.
As an example, a vortex of height field $\varphi_1$ is shown in
Fig. \ref{core3}, one can see that the core of this vortex is a
$(1,1,-1)$ defect. When the vortices of the height fields
condense, which means the height fields are disordered, the system
enters a superfluid phase. When $D$ and $J_{xy}$ are small, the
phase transition occurs when the hopping energy of the defects is
comparable with the gap, the phase boundary is roughly $D \sim
J_{xy}$, as shown in the phase diagram Fig. \ref{phasedia}.

Defect $(1,1,-1)$ can stay at two sublattices of the honeycomb
lattice, let us denote defect $(1,1,-1)$ at sublattice $A$ as
$\psi_{1A}$, denote defect $(1,1,-1)$ at sublattice $B$ as
$\psi_{1B}$; denote defect $(1,-1,-1)$ at sublattice $A$ as
$\psi_{2A}$, defect $(1,-1,-1)$ at sublattice $B$ as $\psi_{2B}$.
Herein 4 flavors of defects are defined because these defects have
independent conservation laws instead of just one global $U(1)$
conservation law in the original Hamiltonian. If we want to hop
one $(1,1,-1)$ defect from sublattice $A$ to sublattice $B$,
global spin configurations within the low energy subspace should
be changed; this means that any local operator cannot hop
$(1,1,-1)$ defect from sublattice $A$ to $B$, i.e. defects at
sublattice $A$ and $B$ are separately conserved. This situation is
similar to the doped quantum dimer model on square lattice
\cite{balents2005b}, in that case the doped holes can also only
hop in one sublattice due to the gauge symmetry of the dimer
model. The gauge symmetry of the dimer model is due to the dimer
constraint imposed automatically.

According to equation (\ref{gaugecharge}), $\psi_{1A}$
($\psi_{1B}$) carries charge $+1$ (-1) of gauge field $a_{1\mu}$,
and $\psi_{2A}$ ($\psi_{2B}$) carries charge $+1$ (-1) of gauge
field $a_{2\mu}$. Now the effective Lagrangian describing the
system is \beqn L = -t|(\partial_\mu - ia_{1\mu})\psi_{1A}|^2
-t|(\partial_\mu + ia_{1\mu})\psi_{1B}|^2 \cr\cr - t|(\partial_\mu
- ia_{2\mu})\psi_{2A}|^2 -t|(\partial_\mu + ia_{2\mu})\psi_{2B}|^2
\cr\cr + \cdots\cdots. \label{fixedpoint}\eeqn

The ellipses include the monopoles of gauge fields as well as the
interaction terms between different matter fields. The interaction
has to be consistent with all the internal symmetries of the
system, which is $U(1)_{global}\times U(1)_{gauge}\times
U(1)_{gauge}$. The $U(1)$ gauge symmetries correspond to the two
flavors of gauge fields, and the $U(1)$ global symmetry
corresponds to the conservation of $S^z$. The regular terms like
$- r|\psi_{ia}|^2 + O(|\psi|^4)$ are all allowed, besides these
terms, another term should in principle exist, which is $-g(
\psi_{1A}\psi_{1B}\psi_{2A}\psi_{2B} + H.c. )$. Four different
flavors of particles can be created and annihilated together,
without any global reconfigurations.

The superfluid phase can be viewed as the condensate of 4 flavors
of matter fields. Let us denote $\psi_{ia} \sim
\exp(\theta_{ia})$, the action can be written as \beqn L =
-\tilde{t}(\partial_\mu \theta_{1A} - a_{1\mu})^2
-\tilde{t}(\partial_\mu \theta_{1B} + a_{1\mu})^2 \cr\cr -
\tilde{t}(\partial_\mu \theta_{2A} - a_{2\mu})^2 -
\tilde{t}(\partial_\mu \theta_{2B} + a_{2\mu})^2 \cr\cr +
g\cos(\theta_{1A} + \theta_{1B} + \theta_{2A} + \theta_{2B}).
\label{6-sf}\eeqn If there is no gauge field, the condensation of
$\theta$s would lead to four gapless Goldstone modes. However, in
the condensate, mode $\theta_{1A} + \theta_{1B} + \theta_{2A} +
\theta_{2B}$ is gapped out by the $g$ term in (\ref{6-sf}), this
implies that in the superfluid phase $\theta_{1A} + \theta_{1B} =
- ( \theta_{2A} + \theta_{2B})$. Meanwhile, $\theta_{1A} -
\theta_{1B}$ will gap out $a_{1\mu}$ through the Higgs mechanism,
$\theta_{2A} - \theta_{2B}$ will also gap out $a_{2\mu} $ through
the Higgs mechanism, therefore the only gapless mode in the
condensate is $\theta_{1A} + \theta_{1B}$ = $ - (\theta_{2A} +
\theta_{2B})$.

Notice that $S^\dagger_i$ can create a pair of $\psi_{1A}$ and
$\psi_{1B}$ particles and also can annihilate a pair of
$\psi_{2A}$ and $\psi_{2B}$ particles.  Therefore we can identify
\beqn S^\dagger \sim \exp[i(\theta_{1A} + \theta_{1B})] = \exp[-
i(\theta_{2A} + \theta_{2B})]. \eeqn Thus, the Goldstone mode
$\theta_{1A} + \theta_{1B}$ is exactly the global $U(1)$ phason
mode of $S^\dagger \sim \exp(i\theta)$.

If one approaches the transition from the superfluid phase, the
transition can be viewed as condensation of vortices in the
superfluid. There are four components of vortices, corresponding
to the four flavors of matter fields. The gapless Goldstone mode
becomes the noncompact $U(1)$ gauge field in the dual language.
The vertex operators in the height field language are the vortex
tunnelling terms. The vertex operators create or annihilate gauge
flux of the original gauge fields $a_{1\mu}$ and $a_{2\mu}$. For
instance, $\exp(2\pi i\varphi_1)$ creates one unit flux of
$a_{1\mu}$. As pointed out in references
\cite{balents2005b,balents2005a}, when one flavor of gauge field
is coupled to two different matter fields, the vortex of each
matter field carries half flux quantum. Since vortex $v_{1A}$ and
$v_{1B}$ carry opposite gauge flux, the vertex operator
$\cos(6\pi\varphi_1)$ corresponds to tunnelling process
$v_{1A}^{\dagger 3}v_{1B}^3 + H.c.$. The dual Lagrangian can be
effectively written as \beqn L = -t|(\partial_\mu -
iA_{\mu})v_{1A}|^2 -t|(\partial_\mu - iA_{\mu})v_{1B}|^2 \cr\cr
-t|(\partial_\mu + iA_{\mu})v_{2A}|^2 -t|(\partial_\mu +
iA_{\mu})v_{2B}|^2 \cr\cr + g(v_{1A}^{\dagger 3}v_{1B}^3 + H.c.) +
g(v_{2A}^{\dagger 3}v_{2B}^3 + H.c.) \cr\cr + g_2
(v^\dagger_{1A}v_{1B}v^\dagger_{2B}v_{2A} + H.c.) + \eta
(v_{1A}v_{1B}v_{2B}v_{2A} + H.c.) \cr\cr + g_1
(v^{\dagger}_{1A}v_{1B}v^{\dagger 2}_{2A}v^2_{2B} + H.c.) +
g_1(v^{\dagger}_{2A}v_{2B}v^{\dagger 2}_{1A}v^2_{1B} + H.c.)
\cr\label{superdual}\eeqn

$A_{\mu}$ in (\ref{superdual}) is the dual form of the Goldstone
``phason" mode in the superfluid phase. $g$, $g_1$ and $g_2$ are
the tunnelling terms due to the vertex operators in
(\ref{vertex2}). Tunnelling term $\eta$ is independent of
monopoles, as this term conserves the total vorticity (consistent
with the $U(1)$ gauge symmetry of (\ref{superdual})), and also
conserves the total gauge flux of the gauge fields $a_{1\mu}$ and
$a_{2\mu}$, therefore it should exist in the field theory.

Let us denote $v_{ia} \sim \exp(-i\chi_{ia})$. After the
condensation of vortices, modes $\chi_{1A} - \chi_{1B}$ and
$\chi_{2A} - \chi_{2B}$ are gapped out by the monopoles. Mode
$\chi_{1A} + \chi_{1B} + \chi_{2A} + \chi_{2B}$ are gapped out by
the $\gamma$ term in equation (\ref{superdual}), i.e. $\chi_{1A} +
\chi_{1B} = - (\chi_{2A} + \chi_{2B})$; also $\chi_{1A} +
\chi_{1B}$ is gapped out by $A_{\mu}$ through the Higgs mechanism.
Therefore in the condensate of vortices, there is no gapless
excitations, which is consistent with the crystalline phase.

If the monopole effect is turned off, the transition point is
described by two gapless noncompact gauge fields and four flavors
of matter fields. However, whether these gapless gauge fields and
matter fields can survive when the monopoles are turned on is an
open question. If the monopoles gap out the gauge field and
confine the matter fields, at the transition there is no gapless
excitation. In this case our theory predicts a direct first order
transition.

The superfluid phase and the plaquette phase break different
symmetries, and according to the classic Landau phase transition
theory, the transition should be either first order, or split into
two transitions, with a disordered phase (or a phase with both
orders) in between. There is no universal law that guarantees one
direct first order transition.

In our theory, the intermediate phases can be understood as the
condensate of composites of defect $\psi_{ia}$. A gapped
disordered phase can be obtained if composites which only carry
local gauge charges but no global $U(1)$ charge are condensed. For
instance, if composites $\psi_{1A}^\dagger\psi_{1B}$ and
$\psi_{2A}^\dagger\psi_{2B}$ are condensed while all the other
composites are disordered, the gauge fields are gapped through the
Higgs mechanism, therefore the height fields are disordered, the
crystalline order disappears. Also, since the composites carry
zero global $U(1)$ charge, there is no gapless Goldstone mode.
Thus we can conclude that the condensate of
$\psi_{1A}^\dagger\psi_{1B}$ and $\psi_{2A}^\dagger\psi_{2B}$ is a
spin disordered phase i.e. a spin liquid phase. Notice that
$\psi_{1A}^\dagger\psi_{1B}$ ($\psi_{2A}^\dagger\psi_{2B}$)
carries two unit gauge charges of gauge field $a_{1\mu}$
($a_{2\mu}$), therefore the condensate of
$\psi_{1A}^\dagger\psi_{1B}$ and $\psi_{2A}^\dagger\psi_{2B}$ is a
spin liquid with $\mathbb{Z}_2 \times \mathbb{Z}_2$ gauge
symmetry, which is the residual gauge symmetry after the
condensation of the composites of matter fields. On the other
hand, if composites carrying only global $U(1)$ charge are
condensed, the superfluid order should coexist with the
crystalline order. For instance, the composite
$\psi_{1A}^\dagger\psi_{1B}^\dagger\psi_{2A}\psi_{2B}$ does not
carry any gauge charge, the condensate of this composite is a
superfluid order, and the crystalline order still exists. Thus
this phase is a supersolid phase.

\section{transition to the nematic phase}

The existence of nematic phase can be derived easily at the
negative large $D$ limit. When $D$ is negative and becomes the
dominant term in the Hamiltonian (\ref{hs1}), the system is
effectively a spin-1/2 system, since $S^z$ on each site can only
be $\pm 1$. The classical ground state of this model is that every
unit triangle should have either $(1,1,-1)$ or $(1,-1,-1)$
configuration. This classical ground state is the same as the
classical Ising model on the kagome lattice, with large
degeneracy. If the same Boltzman weight is imposed for each
classical ground state, the kagome lattice Ising model is
disordered, and the correlation length is finite
\cite{sondhi2001}. By contrast, a related classical system is the
classical Ising model on the triangular lattice, while if the same
Boltzman weight is imposed for each classical ground state, the
triangular lattice Ising model is critical, and an infinitesimal
quantum perturbation is relevant at this critical point and drive
the system into a crystalline phase. If infinitesimal transverse
magnetic field $h\sigma^x$ is turned on, the system is driven to
$\sqrt{3}\times\sqrt{3}$ order \cite{sondhi2001}; if ferromagnetic
$XY$ exchange $-J_{xy}(\sigma^x_{i}\sigma^x_{j} +
\sigma^y_{i}\sigma^y_{j})$ is turned on, the system is driven into
a supersolid phase, which breaks both $U(1)$ symmetry ( $\langle
\sigma^+ \rangle \neq 0$ ), and translational symmetry
\cite{ashvin2005}. Unlike the Ising model on the triangular
lattice, the classical Ising model on the kagome lattice is
disordered, with finite correlation length. In the original
Hamiltonian (\ref{hs1}), if $|J_{xy}| \ll D$, the second order
perturbation generates a term which flips $S^z = 1$ state to $S^z
= -1$ state and vice versa. The effective Hamiltonian reads \beqn
H_{eff} = \sum_{<i,j>} -t(\sigma^x_i\sigma^x_j +
\sigma^y_i\sigma^y_j) + J_z\sigma^z_i\sigma^z_j.
\label{nematic}\eeqn $t \sim J_{xy}^2/D$. As studied in
\cite{ashvin2005,senthil2006}, for the spin-1/2 system on the
kagome lattice, infinitesimal ferromagnetic $XY$ exchange yields
superfluid order, $\langle\sigma^\dagger\rangle \neq 0$. The
spin-1/2 raising operator is the nematic order parameter
$(S^{+})^2$, therefore infinitesimal $|J_{xy}|$ drives the system
into a nematic phase.

Although the nematic phase and the plaquette phase do not
necessarily touch each other in the phase diagram, a direct
transition between these two phases is possible when they are
adjacent in the phase diagram. It is conceivable that a certain
type of spin Hamiltonian can realize the direct transition between
the nematic phase and the six-fold plaquette phase. This direct
transition is more likely to occur when $J_{xy}
> 0$ than the case with $J_{xy} < 0$. In the case with $J_{xy} > 0$, every
hexagon is effectively penetrated by one $\pi$ flux of $a_{1\mu}$
and one $\pi$ flux of $a_{2\mu}$. The motion of defects is
strongly affected by the background magnetic fields, and several
interesting possibilities can happen. One of the possibilities is
that the defects condense in pairs, i.e. $\langle (\psi_{ia})^2
\rangle \neq 0$, as a pair of defects does not see any background
flux. After the pair condensation, the Goldstone mode is
$2(\theta_{1A} + \theta_{1B})$, corresponding to the phase of
$(S^\dagger)^2$, so the system is in the nematic phase discussed
above \cite{senthil2006}. One important difference between the
nematic phase and the superfluid phase is that, each vortex in the
nematic phase only carries one quarter flux of the gauge fields,
therefore the vertex terms in (\ref{vertex2}) correspond to even
higher order of vortex tunnelling processes.

A direct transition between the nematic phase and the plaquette
phase can be described by the following action of paired matter
field $\Psi_{ia} = (\psi_{ia})^2$ \beqn L = -t|(\partial_\mu -
2ia_{1\mu})\Psi_{1A}|^2 -t|(\partial_\mu + 2ia_{1\mu})\Psi_{1B}|^2
\cr\cr - t|(\partial_\mu - 2ia_{2\mu})\Psi_{2A}|^2
-t|(\partial_\mu + 2ia_{2\mu})\Psi_{2B}|^2 \cr\cr - H_I +
\cdots\cdots. \label{fixedpoint2}\eeqn Again the ellipses include
the monopole terms, and $H_I$ contains all the possible
interaction terms between matter fields. Just like the four-defect
creation term discussed in the previous section,
$g\Psi_{1A}\Psi_{1B}\Psi_{2A}\Psi_{2B} + H.c.$ should in principle
exist in the interaction, thus phason mode $\sum_{i = 1}^2\sum_{a
= A}^B\theta_{ia}$ is gapped out in the condensate of $\Psi_{ia}$.
Without the monopole terms this transition is a gapless second
order transition.

Now the question boils down to if the monopole effect is going to
be relevant at the critical point described by action
(\ref{fixedpoint2}). Since the nematic phase is a pair condensate,
each single vortex in the nematic phase carries only one quarter
flux of each flavor of gauge fields, so the vertex operator in
(\ref{vertex2}) corresponds to even higher order of tunnelling
processes than the superfluid case. The dual action now reads

\beqn L = -t|(\partial_\mu - iA_{\mu})v_{1A}|^2 -t|(\partial_\mu -
iA_{\mu})v_{1B}|^2 \cr\cr -t|(\partial_\mu + iA_{\mu})v_{2A}|^2
-t|(\partial_\mu + iA_{\mu})v_{2B}|^2 \cr \cr\cr+
g(v_{1A}^{\dagger 6}v_{1B}^6 + H.c.) + g(v_{2A}^{\dagger
6}v_{2B}^6 + H.c.) \cr\cr + g^\prime (v^{\dagger
2}_{1A}v^2_{1B}v^{\dagger 2}_{2B}v^2_{2A} + H.c.) +
\eta(v_{1A}v_{1B}v_{2B}v_{2A} + H.c.)\cr\cr g_1 (v^{\dagger
2}_{1A}v^2_{1B}v^{\dagger 4}_{2A}v^4_{2B} + H.c.) + g_1(v^{\dagger
2}_{2A}v^2_{2B}v^{\dagger 4}_{1A}v^4_{1B} + H.c.) \label{nemadual}
\cr \eeqn

$g$, $g_1$ and $g_2$ terms are vortex tunnelling processes
corresponding to the vertex operators in (\ref{vertex2}), notice
that now $v_{1A}$ and $v_{1B}$ ($v_{2A}$ and $v_{2B}$) carry one
quarter unit flux of $a_{1\mu}$ ($a_{2\mu}$) . $\eta$ term is a
tunnelling which does not rely on monopole, as it not only
complies with the $U(1)$ gauge symmetry of the dual action
(\ref{nemadual}), but also conserves the flux numbers of the
original gauge fields $a_{1\mu}$ and $a_{2\mu}$. Following the
similar argument as references \cite{senthil2004,senthil2004a},
the monopole terms (vortex tunnelling terms) are likely (but not
rigorously proved) irrelevant at the transition fixed point. It is
known that at the 3D $XY$ fixed point, $\mathbb{Z}_4$ anisotropy
is irrelevant, while the $\mathbb{Z}_3$ anisotropy as in
(\ref{vertex2}) could be relevant. However in our case, the vertex
operators in (\ref{vertex2}) could be irrelevant at the transition
due to the pairing of gauge charges (\ref{fixedpoint2}). Although
the vertex operators are always relevant at the Gaussian fixed
point of (\ref{heighth}), it could be irrelevant at the
order-disorder transition of height fields. It is expected, as in
the calculation that follows, that the scaling dimension of the
vertex operators is approximately proportional to the number of
flavors of matter fields, and proportional to the square of the
product of electric charge and magnetic charge \cite{sachdev1990}.

We can roughly estimate the scaling dimension of the monopole
operators from a random phase approximation (RPA) calculation.
After integrating out the Gaussian part of the matter fields in
(\ref{fixedpoint2}), an effective action for gauge fields
$a_{1\mu}$ and $a_{2\mu}$ is generated \beqn L = \sum_{\alpha =
1}^2\int
\frac{d^3k}{(2\pi)^3}Nn^2|k|\sigma_0|\vec{a}(k)_{\alpha}|^2 +
\cdots\cdots. \eeqn $N$ is the number of flavors of bosons coupled
to each gauge field, $n$ is the number of gauge charge carried by
each boson. In our case $N = n = 2$. In the dual theory, the
kinetic term for the height field is softened to be $\sim k^3$,
and the monopole energy diverges logarithmically instead of
converging in the infrared limit \cite{kleinert2002,herbut2003}.
The dual height fields now have the action \beqn L = \sum_{\alpha
= 1}^2 \int
\frac{d^3k}{(2\pi)^3}\frac{k^3}{Nn^2\sigma_0}|\varphi_{\alpha}|^2
- H_v. \eeqn From this calculation one can see that the scaling
dimension of vertex operators is proportional to $Nn^2$.

$H_v$ in (\ref{vertex2}) contains three types of terms. The
scaling dimensions for $\cos(6\pi\varphi_\alpha)$ and
$\cos(2\pi\varphi_1 + 4\pi\varphi_2)$ calculated from the RPA
approximation is higher than the $\mathbb{Z}_4$ anisotropy studied
before \cite{senthil2004,senthil2004a}. The third vertex operator
is $\cos(2\pi(\varphi_1 - \varphi_2))$, the scaling dimension
calculated from RPA is higher than the $\mathbb{Z}_3$ anisotropy
of 3D $XY$ fixed point, also, on the RPA level, the scaling
dimension is equal to the case with $\mathbb{Z}_4$ anisotropy and
$N = 1$ discussed in reference \cite{senthil2004a}, which has been
shown to be irrelevant at the transition between the Higgs phase
and the confined phase. Recently a Monte Carlo simulation has
shown that the transition between the crystalline phase and the
superfluid phase in a bosonic model with 1/3 filling on the kagome
lattice is a very weak first order transition \cite{ybkim2006}, on
the RPA level, the scaling dimension of the monopole in that case
is smaller than the dimension of all the triple vertices and very
close to the scaling dimension of $\cos(2\pi(\varphi_1 -
\varphi_2))$ in our case. Therefore it is possible that the vertex
operators in our problem are irrelevant at the transition between
the nematic phase and the plaquette order. When the vertex
operators are irrelevant, the critical point is a direct gapless
second order transition, with four flavors of deconfined matter
fields, as well as two flavors of noncompact gauge fields.

\section{transition to the $\sqrt{3}\times\sqrt{3}$ phase}

When $J_{xy} > 0$ and much larger than other coefficients, the
state is most likely to be either the $\sqrt{3}\times\sqrt{3}$
order in Fig. \ref{root3}, or the $q = 0$ state in Fig. \ref{q=0}.
From the $1/S$ expansion, this $\sqrt{3}\times\sqrt{3}$ order is
supposed to be the global minimum of all the classical degenerate
ground states of the Heisenberg model on the kagome lattice at the
isotropic point \cite{henley1995}, although the $q = 0$ state has
also been proved to be one of the local minima. Both
$\sqrt{3}\times\sqrt{3}$ and $q = 0$ states are very typical
configurations for spins on kagome, they can be stabilized by the
second or third nearest neighbor interactions. Also, since both
states are coplanar, they are expected to be even better
candidates in the large $J_{xy}$ case.

Since now the defect hopping is frustrated by the background
magnetic flux of gauge field $a_{1\mu}$ and $a_{2\mu}$ through
each hexagon, the phase angle of the defects cannot be uniformly
distributed on the whole lattice. We will see that the
$\sqrt{3}\times\sqrt{3}$ phase can be interpreted as the
condensate of the four flavors of charge fields in the background
gauge fluxes.

\begin{figure}
\includegraphics[width=2.9in]{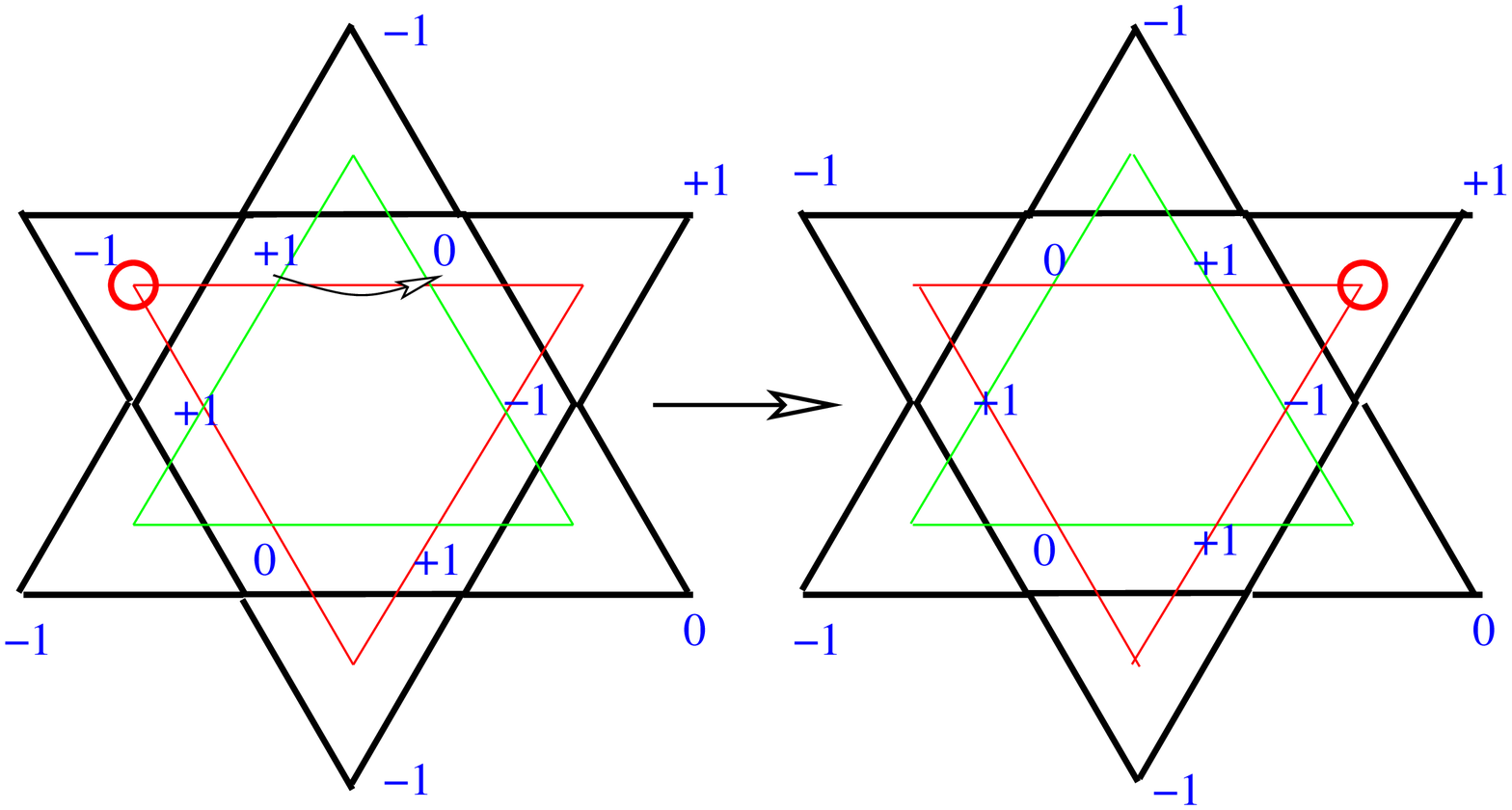}
\caption{Defects which violate the 3-color constraint hop on one
of the two triangular sublattices (red and green) of the honeycomb
lattice.} \label{defecthop}
\end{figure}

Because of the interaction between different matter fields
$g(\psi_{1A}\psi_{1B}\psi_{2A}\psi_{2B} + H.c.)$, we have the
following relation between the phase angles \beqn \theta =
\theta_{1A} + \theta_{1B} = - (\theta_{2A} + \theta_{2B}),\eeqn
$\theta$ is the phase angle of $S^-$. The distribution of phase
$\theta$ can be deduced from the distribution of $\theta_{1A}$ and
$\theta_{1B}$. Notice that $\psi_{1A}$ and $\psi_{1B}$ both live
on the sites of the honeycomb lattice, and hop on two different
triangular sublattices (Fig. \ref{defecthop}). With a background
magnetic field $a_{1\mu}$, the effective Hamiltonian for the
motion of $\psi_{1A}$ is \beqn H = \sum_{<i,j>}t\cos(\theta_{1A,i}
- \theta_{1A,j}) + \cdots\cdots \eeqn This is an antiferromagnetic
$XY$ model on the triangular lattice, and after the condensation
of $\theta_{1A}$, the ground state is the $\sqrt{3}\times\sqrt{3}$
order. This phase can be viewed as the staggered vortex density
phase on the triangular lattice.

\begin{figure}
\includegraphics[width=2.2in]{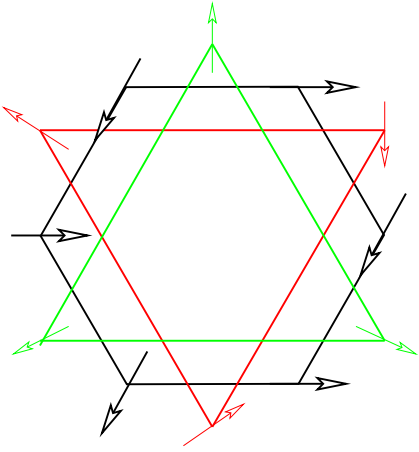}
\caption{The $\sqrt{3}\times\sqrt{3}$ order of $\theta$ obtained
from the ordered pattern of $\theta_{1A}$ and $\theta_{1B}$. The
black arrow corresponds to angle $\theta$, and the pattern of
$\theta$ can be obtained from its adjacent red arrow
($\theta_{1A}$) and green arrow ($\theta_{1B}$).} \label{root32}
\end{figure}

\begin{figure}
\includegraphics[width=2.2in]{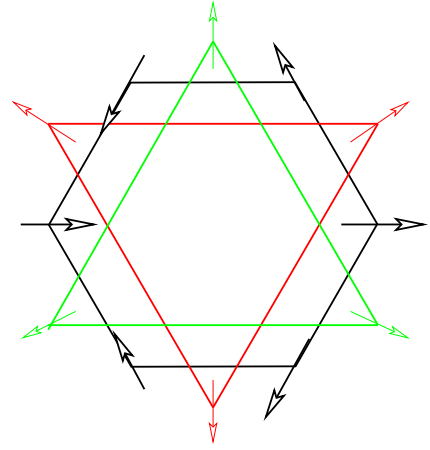}
\caption{$q = 0$ order of $\theta$ obtained from ordered pattern
of $\theta_{1A}$ and $\theta_{1B}$.} \label{q=02}
\end{figure}

If both $\psi_{1A}$ and $\psi_{1B}$ condense (due to the time
reversal symmetry, if $\psi_{1A}$ and $\psi_{1B}$ condense,
$\psi_{2A}$ and $\psi_{2B}$ will also condense), the phase angle
$\theta$ can be determined from the distribution of $\theta_{1A}$
and $\theta_{1B}$. By adding the two ordered patterns of both
$\theta_{1A}$ and $\theta_{1B}$ together, the ordered pattern for
$\theta$ is automatically obtained, and the order can only be
either $q = 0$ state or the $\sqrt{3}\times\sqrt{3}$ state (Fig.
\ref{root32}) and (Fig. \ref{q=02}). In these ordered phases, the
Goldstone mode is still $\theta = \theta_{1A} + \theta_{1B}$.

\section{longitudinal magnetic field}

A longitudinal magnetic field along $\hat{z}$ breaks the time
reversal symmetry in the Hamiltonian, and much of the physics is
significantly changed. Let us assume the magnetic field is small,
i.e. $h S^z$ is much smaller than $J_z$ and $D$ in the model. Note
that this precludes accessing strong-field phenomena such as
magnetization plateaus in this theory.

The six-fold degenerate plaquette phase is expected to survive in
a small longitudinal magnetic field. Since in a small magnetic
field, the classical ground states without $J_{xy}$ are still
configurations with $(1,-1,0)$ triangles only, and therefore the
ring exchange term generated by $J_{xy}$ is still going to select
the plaquette ordered state as the ground state.

However, the excitation energies of defects are changed by the
longitudinal field: $(1,1,-1)$ defects have lower gap than the
$(1,-1,-1)$ defects. Therefore when $|J_{xy}|$ is turned on,
$(1,1,-1)$ defects should condense before $(1,-1,-1)$. As we will
see in this section, the condensate of $(1,1,-1)$ defect is
actually a supersolid phase, with both global $U(1)$ symmetry
breaking and the space symmetry breaking.

Let us take the case with $J_{xy} < 0$ as an example. As long as
$(1,-1,-1)$ defects remain confined and gapped, the total number
of $(1,1,-1)$ defects is conserved before its condensing, because
$(1,1,-1)$ defects cannot be excited without $(1,-1,-1)$ defects
due to the conservation of total $S^z$. Therefore when $(1,1,-1)$
defects condense and $(1,-1,-1)$ defects remain confined and
gapped, the system still has a gapless Goldstone mode due to the
spontaneous breaking of the global conservation of $(1,1,-1)$
defects. The gapless Goldstone mode manifests the superfluid
phase.

Secondly, the spatial symmetry breaking can still be studied in
terms of the dual height fields. Because $(1,1,-1) $ is the vortex
of height field $\varphi_1$, the condensate of $(1,1,-1)$ charge
is the disordered phase of $\varphi_1$, $\varphi_1$ no longer has
nonzero expectation values. However, because $\varphi_2$ is still
ordered, the vertex operator $-\alpha\cos(6\pi\varphi_2)$ in
(\ref{vertex2}) has 3 minima, corresponding to 3 fold degenerate
states. These 3 minima are the plaquette orders of
$(-1,0,-1,0,-1,0)$ hexagon on 3 different sublattices.

The height fields $\varphi_1$ and $\varphi_2$ are coupled through
the vertex operators as shown in (\ref{vertex2}). This coupling is
not going to lift the 3 fold degeneracy of $\varphi_2$ when
$\varphi_1$ is disordered. The reason is as follows: The whole
action (\ref{heighth}) is invariant under transformation
$\varphi_1 \rightarrow \varphi_1 + 1/3$, $\varphi_2 \rightarrow
\varphi_2 + 1/3$. Since $\varphi_1$ is disordered, after
integrating over $\varphi_1$, the effective action
$H_{eff}(\varphi_2)$ for $\varphi_2$ does not break this symmetry,
and the leading vertex term generated from integrating out
$\varphi_1$ is $\cos(6\pi\varphi_2)$. This can be clearly seen
from the following equations \beqn \exp(- H_{eff}(\varphi_2)) =
\int D\varphi_1\exp (- H(\varphi_1,\varphi_2)) \cr\cr = \int
D\varphi_1\exp (-H(\varphi_1 + 1/3, \varphi_2 + 1/3)) = \cr\cr
\int D(\varphi_1 + 1/3) \exp(- H(\varphi_1 + 1/3, \varphi_2 +
1/3)) \cr\cr = \int D\varphi_1\exp (- H(\varphi_1, \varphi_2 +
1/3))\cr\cr = \exp(- H_{eff}(\varphi_2 + 1/3)). \eeqn Notice that
the above proof is only valid if $\varphi_1$ does not take any
nonzero expectation value, i.e. $\varphi_1$ is in disordered
phase. Alternatively, one can understand this argument from the
conservation of the gauge fluxes. Vertex operator
$\cos(2\pi\varphi_\alpha)$ can annihilate or create one unit flux
of gauge field $a_{\alpha\mu}$. The total flux of both gauge
fields is conserved mod 3 in vertex operator Hamiltonian
(\ref{vertex2}), and as the disordered phase of $\varphi_1$ (the
condensate of $(1,1,-1)$ defect) does not tend to violate this
conservation, the resultant effective Hamiltonian after
integrating out $\varphi_1$ fields does not break the
$\mathbb{Z}_3$ conservation of total gauge flux, i.e. the lowest
order vertex operator of the resultant effective Hamiltonian of
$\varphi_2$ is $\cos(6\pi \varphi_2)$. Therefore the 3-fold
degenerate plaquette order is not lifted. A similar result is
obtained for $J_{xy} > 0$ too. Thus, the phase with defect
$(1,1,-1)$ condensed while defect $(1,-1,-1)$ confined breaks both
spatial symmetry and the global $U(1)$ symmetry, and hence must be
the supersolid phase.

When $D < 0$ and large, a small longitudinal magnetic field
changes the physics severely. In this regime, the classical ground
state has Ising configuration $(1,1,-1)$ on each triangle. In the
previous sections we mentioned that the classical Ising ground
state on the kagome lattice is disordered with finite correlation
length. However, once the longitudinal magnetic field is turned
on, every triangle has $(1,1,-1)$ configuration. Since the sites
of the kagome lattice are the links of the dual honeycomb lattice,
the ground state configurations with small magnetic field can be
mapped onto the dimer configurations on the honeycomb lattice,
with $S^z = -1$ mapped onto dimer, and $S^z = 1$ mapped onto empty
link. If the same Boltzmann weight is imposed on every dimer
configuration, the system is again critical, with power law
decaying spin-spin correlation function.

Since the classical ground state is critical, it is again very
instable against quantum perturbations. If a small $J_{xy}$ is
turned on, the system is driven into a gapped crystalline phase.
At the sixth-order perturbation, a ring exchange term is generated
\beqn H_{ring} = \sum_{\hexagon} - t((S^\dagger_1S_2)^2
(S^\dagger_3S_4)^2 (S^\dagger_5S_6)^2 + H.c.) \label{ring2}\eeqn
$t\sim J^6_{xy}/(J_z^2D^3)$. This ring exchange term plays the
same role as the dimer flipping term in the honeycomb lattice
quantum dimer model, which will generally lead to a crystalline
phase. Notice that, besides the off-diagonal flipping term in
(\ref{ring2}), diagonal terms are also generated. According to
several previous works \cite{bergman2006a,bergman2006b}, the
diagonal terms generated by perturbation theory favor flippable
hexagons, therefore it is expected that the crystalline order is
either plaquette order or columnar order \cite{sondhi2003a}.

\begin{figure}
\includegraphics[width=2.9in]{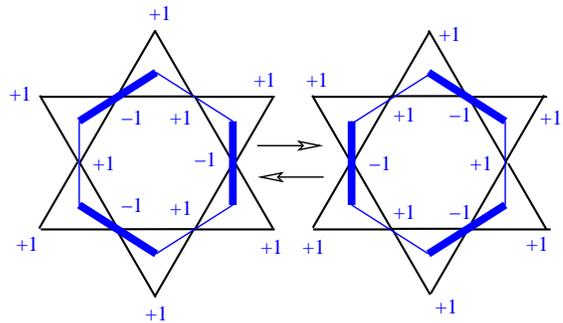}
\caption{The distribution of $S^z = 1$ and $S^z = -1$ sites on the
kagome lattice is equivalent to the distribution of dimers of
dimer model on the dual honeycomb lattice. The ring exchange term
(\ref{ring2}) plays the same role as the dimer resonating term in
the quantum dimer model.} \label{dimerring}
\end{figure}

\begin{figure}
\includegraphics[width=3.0in]{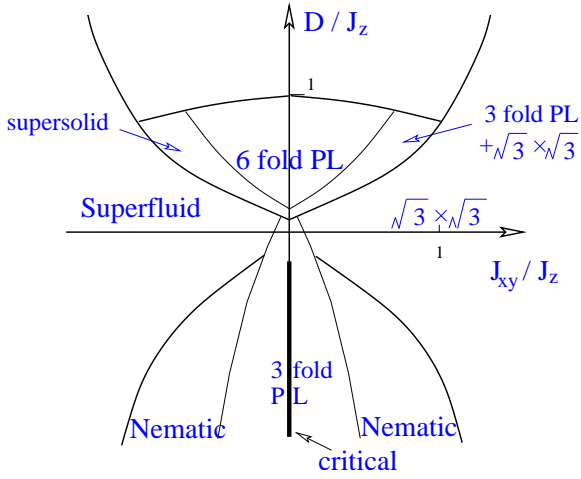}
\caption{The phase diagram in small longitudinal magnetic field
$h$. The difference between this phase diagram and the one without
magnetic field in Fig. \ref{phasedia} is that, between the
plaquette phase and the superfluid phase there is a supersolid
phase. And in the negative and large $D$ case, small $J_{xy}$ is
going to generate a gapped crystalline order.} \label{phasediah}
\end{figure}

Presumably the crystalline phase disappears when $J_{xy}^2/D \sim
h$, Since the sign of $t$ in (\ref{ring2}) is always positive
(independent of the sign of $J_{xy}$), the crystalline phase
should extends symmetrically on the two sides of the classical
line with $J_{xy} = 0$, until the system enters the nematic phase.
The sketchy phase diagram in a small magnetic field is shown in
Fig. \ref{phasediah}, note that this phase diagram involves a lot
of phases, the detailed topology of the phase diagram would depend
on the details of the microscopic model.

\section{the $SU(2)$ point}

At the isotropic point $J_{xy} = J_z$ and $D = 0$, the
$\sqrt{3}\times\sqrt{3}$ state is just one possibility. This state
is obtained from quantum perturbation on the classical limit. If
we start with the quantum limit, another possible state can be
obtained: the $q = 0$ dimer plaquette state.

\begin{figure}
\includegraphics[width=2.9in]{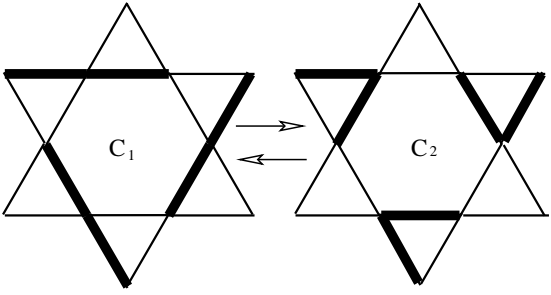}
\caption{The dimer model on the kagome lattice. Since this is a
spin-1 system, every site connects to two dimers. The dimer
flipping term flips configuration $\mathcal{C}_1$ to
$\mathcal{C}_2$ and vice versa.} \label{c1c2}
\end{figure}

This state can be understood quite easily from the quantum dimer
model on the kagome lattice. For a spin-1 system, each site can
form two spin singlets, which means each site connects to two
dimers. Since every site on the kagome lattice is shared by four
links, this dimer model is half filled. The dimer resonance term
on the kagome lattice is shown in Fig. \ref{c1c2}, which can flip
the dimer covering $\mathcal{C}_1$ to $\mathcal{C}_2$ and vice
versa. The dimer model Hamiltonian is now written as \beqn H =
-t(|\mathcal{C}_1 \rangle \langle \mathcal{C}_2| + H.c.) +
V(|\mathcal{C}_1\rangle\langle \mathcal{C}_1| +
|\mathcal{C}_2\rangle\langle \mathcal{C}_2|).\eeqn

As long as $V < t$, the exact ground state wave function of this
Hamiltonian can be written as \beqn |\Psi\rangle = \frac{1}{Z}
\prod_{\hexagon}(| \mathcal{C}_1 \rangle +
|\mathcal{C}_2\rangle).\eeqn This state does not break any space
symmetry, and it minimizes the energy of each hexagon
individually. This state should be the ground state of a certain
type $SU(2)$ invariant spin Hamiltonian.

Now the question is whether this dimer plaquette phase is a new
phase or it can be continuously connected with one of the other
states discussed early this paper without any physical
singularity. Notice that the gapped photon phase with $(0,0,0)$ on
every triangle breaks no space symmetry either, thus one can
imagine adding $D(S^z)^2$ on the isotropic Hamiltonian, and the
ground state wave function can be continuously deformed to the
gapped photon phase.

\begin{figure}
\includegraphics[width=3.2in]{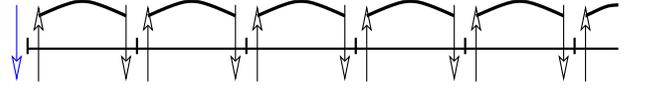}
\caption{A typical state in the Haldane phase. In the bulk, every
site is shared by two dimers. However, at each edge, there is one
residual unpaired spin-1/2 variable.} \label{haldane}
\end{figure}

\begin{figure}
\includegraphics[width=3.0in]{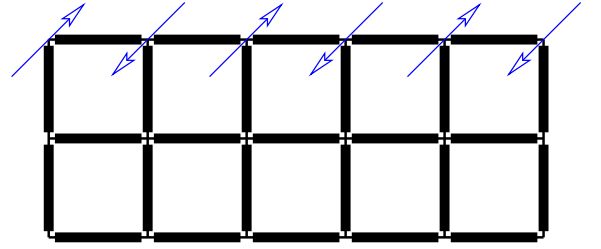}
\caption{The edge of AKLT state on the square lattice. There is
one extra spin-1/2 degree of freedom on each site of the edge,
therefore the edge state is effectively a spin-1/2 spin chain,
which is either gapless or breaks translational symmetry. The AKLT
state is qualitatively different from the state with $S^z = 0$
everywhere.}\label{squareedge}
\end{figure}

\begin{figure}
\includegraphics[width=3.0in]{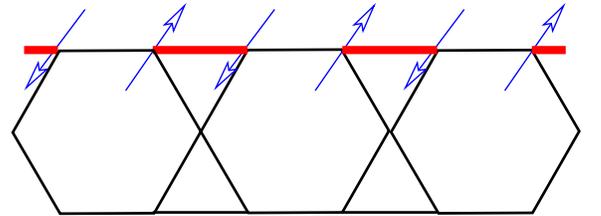}
\caption{The edge of the kagome lattice. Every unit cell at the
edge has even number of spin-1/2 quantities. Presumably the edge
state is gapped and featureless.} \label{kagomeedge}
\end{figure}

One has to be careful about the naive argument above. Let us
consider the one dimensional analogues of the dimer plaquette
phase and the (0,0,0) phase as a check of our naive argument. The
one dimensional Haldane phase \cite{haldane1988} for spin-1
Heisenberg chain is gapped, and breaks no symmetry. One can
imagine that by adding $D(S^z)^2$ on the Heisenberg Hamiltonian
the Haldane phase will be continuously connected to the state with
$S^z = 0$ everywhere. However, the spin-1 Heisenberg chain is
characterized by two special properties: the first is the
existence of gapless edge states, the second is the hidden diluted
antiferromagnetic order. The existence of the gapless edge states
can be understood as follows: All the sites in the bulk are shared
by two dimers, while the site at the edge only connects to one
dimer, hence there is a residual spin-1/2 degree of freedom at
each edge (Fig. \ref{haldane}). The hidden diluted
antiferromagnetic order can be viewed from expanding the AKLT
state (the explicit wavefunction of one particular state in the
Haldane phase) of the spin-1 chain in the basis of $S^z$. In this
expansion, one typical state is as follows \beqn
+-00000+000-+00-+000 \cdots\cdots \eeqn Every $S^z = 1$ site is
always followed by one $S^z = -1$ site, although there could be a
number of $S^z = 0$ sites in between. A special nonlocal string
operator could be introduced to describe this hidden order in the
Haldane phase \cite{nijs1987,nijs1989}.

The nice features of the Haldane phase also exist in the AKLT
state of spin-2 systems on the square lattice. Let us take a
cylinder geometry as an example. There is one unpaired spin-1/2
degree of freedom on each site of the edge, therefore the edge
state is effectively a spin-1/2 chain, which is either gapless or
gapped but breaks translational symmetry. The AKLT state is
qualitatively different from the state with $S^z = 0$ everywhere
as well (Fig. \ref{squareedge}).

However, the edge states are missing in the dimer plaquette phase
of the spin-1 dimer model on the kagome lattice. If we take a
cylinder with edges, on each site of the edge there is also a
residual spin-1/2 degree of freedom. However, due to the geometry
of the kagome lattice, in every unit cell of the edge there are
even number of spins, therefore effectively the edge state is a
chain with integer spin (Fig. \ref{kagomeedge}). The resultant
edge state is generally gapped and featureless at the edge, which
is the Haldane phase on a closed circle. Thus, we conclude that
the dimer plaquette phase can be continuously connected with the
gapped photon phase, with $(0,0,0)$ spin configuration on every
triangle.

\section{Other transitions}

In the phase diagrams Fig. \ref{phasedia} and Fig.
\ref{phasediah}, there are several other transitions which are
interesting. First of all, in Fig. \ref{phasedia}, the transition
between the nematic phase and the superfluid phase is probably an
Ising transition, as this transition breaks the $\mathbb{Z}_2$
symmetry in the nematic state. The transition between the nematic
phase and the $\sqrt{3}\times\sqrt{3}$ phase is supposed to be a
first order transition.

In the case with magnetic field, since another crystalline order
is opened up (Fig. \ref{phasediah}), there is a transition between
the crystalline phase and the nematic phase. However, now that in
the case of large and negative $D$, the system can be described by
an effective spin-1/2 model (\ref{nematic}), the transition can be
understood as the transition between the crystalline order and the
superfluid order of hard core bosons on the kagome lattice. This
transition has been studied in references
\cite{ybkim2006,ybkim2006A}, and the transition is a weak first
order transition.

\section{conclusions and extensions}

In the current work we studied the global phase diagram of the
spin-1 XXZ antiferromagnet on the kagome lattice. Various phases
which have been studied before can be obtained from condensation
of the defects in one single phase. The phase diagram was also
obtained for the case of a magnetic field along the $z$ direction.
One route to test this phase diagram experimentally is by neutron
scattering or other measurements on the spin-1 kagome materials,
for instance $\mathrm{Ni_3V_2O_8}$ and $\mathrm{BF_4}$ salts.

In all the previous sections, the model under consideration only
contains quadratic interactions. However, in some circumstances,
for instance a spin-1 bosonic system trapped in an optical
lattice, the biquadratic interactions
$-J_2(\vec{S}_1\cdot\vec{S}_2)^2$ have been shown to be important
\cite{demler2003}. This biquadratic term can help to stabilize the
nematic phase, when it becomes the dominant term in the
Hamiltonian.  In closing we briefly explain one interesting
consequence of this biquadratic interaction, in case a cold-atom
realization of this Hamiltonian is constructed.

Suppose that the system is in the six-fold degenerate plaquette
ordered state, and let us gradually turn on the biquadratic term
in the XY plane \beqn H_{bi} = - J_2(S^x_iS^x_j + S^y_iS^y_j)^2.
\label{bi}\eeqn This biquadratic term is consistent with the $XXZ$
symmetry of our model. At the third order in perturbation theory,
this biquadratic term can generate resonance between $(+1, -1, +1,
-1, +1, -1)$ hexagon with $(-1, +1, -1, +1, -1, +1)$. Notice that,
although spin $S^z = +1$, $-1$ and $0$ are treated as three
colors, the $\mathbb{Z}_3$ symmetry is missing in Hamiltonian
(\ref{ringgauge}), since the resonances were only between $(+1, 0,
+1, 0, +1, 0)$ hexagons and between the $(-1, 0, -1, 0, -1, 0)$
hexagons.  Therefore the full $\mathbb{Z}_3$ symmetry can be
restored by turning on the biquadratic term (\ref{bi}). At this
$\mathbb{Z}_3$ point the ground state is probably a nine-fold
degenerate plaquette order. Once the biquadratic XY exchange
dominates the quadratic XY exchange, the phase becomes a
three-fold degenerate plaquette ordered state with resonating
$(+1, -1)$ hexagons.

\acknowledgments

The authors thank L. Balents, D. Huse and A. Vishwanath for useful
conversations and NSF DMR-0238760 for support.

\bibliography{kagome}
\end{document}